\documentclass[usenatbib]{mn2e}
\bibpunct{(}{)}{;}{a}{}{,}
\usepackage[dvips]{graphicx}

\newcommand{\mic}{\hbox{$\mu$m}}
\newcommand{\hii}{\mbox{H\,{\sc ii}}}
\newcommand{\hi}{\mbox{H\,{\sc i}}}
\newcommand{\tbgswarm}{\hbox{$T_\mathrm{W}^{\,\mathrm{BC}}$}}
\newcommand{\tbgswarmism}{\hbox{$T_\mathrm{W}^{\,\mathrm{ISM}}$}}
\newcommand{\tbgscold}{\hbox{$T_\mathrm{C}^{\,\mathrm{ISM}}$}}
\newcommand{\fmu}{\hbox{$f_\mu$}}
\newcommand{\ldust}{\hbox{$L_{\mathrm{d}}^{\,\mathrm{tot}}$}}
\newcommand{\ldbc}{\hbox{$L_{\mathrm{d}}^{\,\mathrm{BC}}$}}
\newcommand{\ldism}{\hbox{$L_{\mathrm{d}}^{\,\mathrm{ISM}}$}}
\newcommand{\lseven}{\hbox{$L_\nu^{\mathrm{6.75}}$}}
\newcommand{\leight}{\hbox{$L_\nu^{\mathrm{8}}$}}
\newcommand{\ltwelve}{\hbox{$L_\nu^{\mathrm{12}}$}}
\newcommand{\lfifteen}{\hbox{$L_\nu^{\mathrm{15}}$}}
\newcommand{\ltwofour}{\hbox{$L_\nu^{\mathrm{24}}$}}
\newcommand{\ltwofive}{\hbox{$L_\nu^{\mathrm{25}}$}}
\newcommand{\lsixty}{\hbox{$L_\nu^{\mathrm{60}}$}}
\newcommand{\lseventy}{\hbox{$L_\nu^{\mathrm{70}}$}}
\newcommand{\lhundred}{\hbox{$L_\nu^{\mathrm{100}}$}}
\newcommand{\lonesixty}{\hbox{$L_\nu^{\mathrm{160}}$}}
\newcommand{\fseven}{\hbox{$F_\nu^{\mathrm{6.75}}$}}
\newcommand{\feight}{\hbox{$F_\nu^{\mathrm{8}}$}}
\newcommand{\ftwelve}{\hbox{$F_\nu^{\mathrm{12}}$}}
\newcommand{\ffifteen}{\hbox{$F_\nu^{\mathrm{15}}$}}
\newcommand{\ftwofour}{\hbox{$F_\nu^{\mathrm{24}}$}}
\newcommand{\ftwofive}{\hbox{$F_\nu^{\mathrm{25}}$}}
\newcommand{\fsixty}{\hbox{$F_\nu^{\mathrm{60}}$}}
\newcommand{\fseventy}{\hbox{$F_\nu^{\mathrm{70}}$}}
\newcommand{\fhundred}{\hbox{$F_\nu^{\mathrm{100}}$}}
\newcommand{\fonesixty}{\hbox{$F_\nu^{\mathrm{160}}$}}
\newcommand{\xipahs}{\hbox{$\xi_\mathrm{PAH}^\mathrm{\,BC}$}}
\newcommand{\xipahsism}{\hbox{$\xi_\mathrm{PAH}^\mathrm{\,ISM}$}}
\newcommand{\xipahstot}{\hbox{$\xi_\mathrm{PAH}^\mathrm{\,tot}$}}
\newcommand{\ximir}{\hbox{$\xi_\mathrm{MIR}^\mathrm{\,BC}$}}
\newcommand{\ximirism}{\hbox{$\xi_\mathrm{MIR}^\mathrm{\,ISM}$}}
\newcommand{\ximirtot}{\hbox{$\xi_\mathrm{MIR}^\mathrm{\,tot}$}}
\newcommand{\xiwarm}{\hbox{$\xi_\mathrm{W}^\mathrm{\,BC}$}}
\newcommand{\xiwarmism}{\hbox{$\xi_\mathrm{W}^\mathrm{\,ISM}$}}
\newcommand{\xiwarmtot}{\hbox{$\xi_\mathrm{W}^\mathrm{\,tot}$}}
\newcommand{\xicold}{\hbox{$\xi_\mathrm{C}^\mathrm{\,ISM}$}}
\newcommand{\xicoldtot}{\hbox{$\xi_\mathrm{C}^\mathrm{\,tot}$}}

\newcommand{\xibgsbc}{\hbox{$\xi_\mathrm{BG}^\mathrm{\,BC}$}}
\newcommand{\mwarmbc}{\hbox{$M_\mathrm{W}^\mathrm{\,BC}$}}
\newcommand{\mwarmism}{\hbox{$M_\mathrm{W}^\mathrm{\,ISM}$}}
\newcommand{\mcoldism}{\hbox{$M_\mathrm{C}^\mathrm{\,ISM}$}}
\newcommand{\tg}{\hbox{$t_{\mathrm{g}}$}}
\newcommand{\tauv}{\hbox{$\hat{\tau}_{V}$}}
\newcommand{\ha}{\hbox{H$\alpha$}}
\newcommand{\hb}{\hbox{H$\beta$}}
\newcommand{\ssfr}{\hbox{$\psi_{\mathrm S}$}}
\newcommand{\mdms}{\hbox{$M_{\mathrm d}/M_\ast$}}
\newcommand{\mdmg}{\hbox{$M_{\mathrm d}/M_{\mathrm H}$}}

\begin{document}

\title[A simple model to interpret the emission from galaxies]{A simple
model to interpret the ultraviolet, optical and infrared emission from galaxies}
\author[E. da Cunha et al.]{Elisabete~da~Cunha$^{1}$\thanks{E-mail:
dacunha@iap.fr}, 
St\'ephane~Charlot$^{1}$ and David~Elbaz$^{2}$ \\
$^{1}$Institut d'Astrophysique de Paris, CNRS, Universit\'e Pierre \&
Marie Curie, 98 bis Boulevard Arago, 75014 Paris, France \\
$^{2}$Laboratoire AIM, CEA/DSM-CNRS-Universit\'e Paris Diderot,
DAPNIA/Service d'Astrophysique, \\
CEA Saclay, Orme des Merisiers, 91191 Gif-sur-Yvette Cedex, France}
\date{Accepted 2008 June 04. Received 2008 May 30; in original form 2007 December 18}

\maketitle
\begin{abstract}

We present a simple, largely empirical but physically motivated model
to interpret the mid- and far-infrared spectral energy distributions of
galaxies consistently with the emission at ultraviolet, optical and 
near-infrared wavelengths.
Our model relies on an existing angle-averaged prescription to compute
the absorption of starlight by dust in stellar birth clouds and in the 
ambient interstellar medium (ISM) in galaxies. 
We compute the spectral energy distribution of the power reradiated
by dust in stellar birth clouds as the sum of three components:
a component of polycyclic aromatic hydrocarbons (PAHs); a mid-infrared
continuum characterising the emission from hot grains at temperatures
in the range 130--250~K; and a component of grains in thermal 
equilibrium with adjustable temperature in the range 30--60~K. In 
the ambient ISM, we fix for simplicity the relative proportions of these
three components to reproduce the spectral shape of diffuse cirrus 
emission in the Milky Way, and we include a component of cold grains 
in thermal equilibrium with adjustable temperature in the range 15--25~K.
Our model is both simple and versatile enough that it can be used to 
derive statistical constraints on the star formation histories and dust
contents of large samples of galaxies using a wide range of ultraviolet, 
optical and infrared observations.
We illustrate this by deriving median-likelihood estimates of the 
star formation rates, stellar masses, effective dust optical depths,
dust masses, 
and relative strengths of different dust components of 66 well-studied
nearby star-forming galaxies from the Spitzer Infrared Nearby Galaxy 
Survey (SINGS). We explore how the constraints derived in this way 
depend on the available spectral information.
From our analysis of the SINGS sample, we conclude that the mid- and 
far-infrared colours of galaxies correlate strongly with the specific
star formation rate, as well as with other galaxy-wide quantities 
connected to this parameter, such as the ratio of infrared luminosity 
between stellar birth clouds and the ambient ISM, the contributions by 
PAHs and grains in thermal equilibrium to the total infrared
emission, and the ratio of dust mass to stellar mass. 
Our model can be straightforwardly applied to interpret ultraviolet,
optical and infrared spectral energy distributions from any galaxy sample.

\end{abstract}

\begin{keywords}
dust, extinction -- galaxies: ISM, stellar content -- infrared: galaxies.
\end{keywords}


\section{Introduction}
\label{intro}

The spectral energy distributions of galaxies contain valuable 
information about their contents in stars, gas and dust. Direct 
ultraviolet, optical and near-infrared radiation from stars provides
clues on the past star formation history, chemical enrichment and
attenuation by dust. Nebular emission lines produced by the gas
heated by young stars provide further clues on the current star formation 
activity and the physical conditions of the star-forming gas. At 
wavelengths $\lambda\ga 3~\mu$m, the mid- and far-infrared emission 
reflects the heating of dust in different components of the interstellar
medium (ISM) by stars of all ages. Observations at ultraviolet, optical
and infrared wavelengths are now becoming available for large samples
of galaxies. These include data collected in the ultraviolet by the
{\it Galaxy Evolution Explorer} ({\it GALEX}, \citealt{MARTIN05}), in
the optical by the Two-degree Field Galaxy Redshift Survey \citep{2dF}
and the  Sloan Digital Sky Survey (SDSS, \citealt{S02}), in the 
near-infrared by the Two Micron All Sky Survey (2MASS, 
\citealt{2MASS}), in the mid- and far-infrared by the {\it Infrared
Astronomical Satellite} ({\it IRAS}, \citealt{IRAS}), the {\it Infrared
Space Observatory} ({\it ISO}, \citealt{ISO}) and the {\it Spitzer 
Space Telescope} \citep{SIRTF}, and in the sub-millimetre by the
Sub-millimeter Common User Bolometer Array (SCUBA) on the James 
Clerk Maxwell Telescope \citep{SCUBA}. Extracting constraints on the
stellar populations and ISM of galaxies from these multi-wavelength
observations requires the consistent modelling of the emission by stars,
gas and dust.

A standard approach to model consistently the emission from 
stars and dust in galaxies has been to solve the radiative transfer 
equation for idealised (bulge + disc) spatial distributions of 
stars and dust 
(e.g. \citealt{RR80,ERR90,GORDON01,MIS01,POPESCU00,MISIRIOTIS01}). 
Early models of this type did not include the evolution of stellar
populations. \citet{GRASIL98} were the first to couple radiative 
transfer through a dusty ISM and the spectral (and even chemical) 
evolution of stellar populations. Their model also accounts for the
fact that stars are born in dense molecular clouds, which dissipate 
after some time, and hence, that newly born stars are more attenuated
than older stars (see also, e.g., \citealt{CF00,TUFFS04}).
This type of sophisticated model is useful to 
interpret in detail the emission from individual galaxies in terms
of constraints on stellar populations and the spatial distribution 
and physical properties of the dust. However, because of the
complexity of radiative transfer computations, it is not optimised to
derive statistical constraints from observations of large samples
of galaxies.

A more recent model of starburst galaxies by \citet[][see also 
\citealt{GROVES07}]{DOPITA05} incorporates the consistent treatment
of the spectral evolution of stellar populations, the dynamic expansion
of \hii\ regions and radiative transfer of starlight through gas
and dust. The authors of this model provide a simple parameterization 
of the ultraviolet, optical and infrared spectra of starburst galaxies 
by adding the spectra of different types of compact \hii\ regions 
and their surrounding photo-dissociation regions. This model provides
a fast and flexible tool to interpret starburst galaxy spectra in 
terms of the physical parameters of star-forming regions. However,
it is not designed to be applicable to more quiescent galaxies, in 
which older stellar populations dominate the emission.

In parallel to these theoretical studies, observations by {\it IRAS} 
and {\it ISO} have motivated the development of simple, empirically
calibrated spectral libraries to interpret the infrared emission
from galaxies at wavelengths between 3 and 1000~\mic. For example,
\cite{CE01} and \cite{DH02} both present single-parameter families of 
infrared spectra to relate an observed spectral energy distribution 
to either the total infrared luminosity of a galaxy or the intensity
of the interstellar radiation field heating the dust. These libraries
can be applied easily to the interpretation of large galaxy samples.
They have proved useful to characterise the infrared emission from
observed galaxies and to investigate the origin of the cosmic infrared
background (e.g. \citealt{FRANC01, CE01,DE02,LAGACHE03,LAGACHE04,
DALE05,MARC06}). 
A disadvantage of this approach is that it does not relate consistently
the infrared emission of the dust to the emission from stellar 
populations. Another potential limitation is that most existing
spectral libraries were calibrated using local galaxy samples, and 
hence, they may not be applicable to studies of the infrared emission
from galaxies at all redshifts (e.g. \citealt{POPE06, ZHENG07}).

In this paper, we present a simple, largely empirical but physically
motivated model to interpret the mid- and far-infrared spectral
energy distributions of galaxies consistently with the emission 
at ultraviolet, optical and near-infrared wavelengths. We compute 
the spectral evolution of stellar populations using the \cite{BC03} 
population synthesis code. To describe the attenuation of starlight
by dust, we appeal to the two-component model of \cite{CF00}. This
has been shown to account for the observed relations between the 
ultraviolet and optical (line and continuum) spectra and the {\em 
total} infrared luminosities of galaxies in wide ranges of 
star-formation activity and dust content \citep{JB04,KONG04}. We
use this model to compute the luminosity absorbed and re-emitted by
dust in stellar birth clouds (i.e. giant molecular clouds) and in 
the ambient (i.e. diffuse) ISM in galaxies. We then distribute this
luminosity in wavelength to compute infrared {\em spectral energy 
distributions}. We describe the infrared emission from stellar birth
clouds as the sum of three components: a component of polycyclic 
aromatic hydrocarbons (PAHs); a mid-infrared continuum characterising
the emission from hot grains at temperatures in the range 130--250~K;
and a component of grains in thermal equilibrium with adjustable 
temperature in the range 30--60~K. In the ambient ISM, we fix for
simplicity the relative proportions of these three components to 
reproduce the spectral shape of diffuse cirrus emission in the Milky
Way, and we include a component of cold grains in thermal equilibrium 
with adjustable temperature in the range 15--25~K.

This simple but versatile model allows us to derive statistical 
estimates of physical parameters such as star formation rate, stellar
mass, dust content and dust properties, from combined ultraviolet, 
optical and infrared galaxy spectra. To achieve this, we adopt a 
Bayesian approach similar to that successfully employed by, e.g.,
\citet{KAUF03}, \citet{JB04}, \citet{AG05} and \citet{SAL07} to 
interpret ultraviolet, optical and near-infrared galaxy spectra 
using only the \cite{BC03} and \cite{CF00} models. As an example,
we derive median-likelihood estimates of a set of physical 
parameters describing the stellar and dust contents of 66 
star-forming galaxies from the Spitzer Infrared Nearby Galaxy 
Survey (SINGS, \citealt{K03}). Our model reproduces well the observed
spectral energy distributions of these galaxies across the entire 
wavelength range from the far-ultraviolet to the far-infrared, 
and the star formation histories and dust contents of the galaxies
are well constrained. We explore how the constraints derived in 
this way depend on the available spectral information. From our 
analysis of the SINGS sample, we conclude that the mid- and 
far-infrared colours of galaxies are tightly related to the 
specific star formation rate and to other galaxy-wide properties
connected to this parameter.

We present our model of the combined ultraviolet, optical and infrared 
spectral energy distributions of galaxies in Section~\ref{the_model}.
In Section~\ref{extraction}, we first describe our approach to derive 
statistical constraints on galaxy physical parameters from multi-wavelength 
observations. Then, we use this approach to interpret observations of the
SINGS galaxy sample taken with {\it GALEX}, 2MASS, {\it Spitzer}, {\it
ISO}, {\it IRAS} and SCUBA.
We compare our results to those that would be 
obtained using previous prescriptions of the infrared emission from 
galaxies. We also discuss potential sources of systematic errors. 
Section~\ref{conclusion} summarises our conclusions.


\section{The model}
\label{the_model}
	
In this section, we describe our model to compute the mid- and 
far-infrared spectral energy distributions of galaxies consistently
with the emission at ultraviolet, optical and near-infrared wavelengths.
In Section~\ref{stellar}, we first briefly review the stellar population
synthesis code and the two-component dust model used to compute the 
emission of starlight and its transfer through the ISM in galaxies. Then,
in Section~\ref{irmodel}, we present our model to compute the spectral 
energy distribution of the infrared emission from dust. We calibrate this
model using a sample of 107 nearby star-forming galaxies observed by {\it
IRAS} and {\it ISO}. In Section~\ref{combination}, we show examples of
combined ultraviolet, optical and infrared spectral energy distributions
of different types of star-forming galaxies.


	\subsection{Stellar emission and attenuation by dust}
	\label{stellar}

We use the latest version of the \cite{BC03} stellar population
synthesis code to compute the light produced by stars in galaxies. This
code predicts the spectral evolution of stellar populations at 
wavelengths from 91 \AA~to 160~\mic\ and at ages between $1\times 
10^{5}$ and $2\times 10 ^{10}$ years, for different metallicities, 
initial mass functions (IMFs) and star formation histories. 
We use the most recent version of the
code, which incorporates a new prescription by \cite{MG07} for the TP-AGB 
evolution of low- and  intermediate-mass stars. The main effect of this
prescription is to improve the predicted near-infrared colours of 
intermediate-age stellar populations (\citealt{BRUZ07}; see also 
Charlot \& Bruzual, in preparation).
In all applications throughout this paper, we adopt the Galactic-disc IMF 
of \citet{CHAB03}.

We compute the attenuation of starlight by dust using the simple, 
angle-averaged model of \cite{CF00}. This accounts for the fact that
stars are born in dense molecular clouds, which dissipate typically on
a timescale $t_0 \sim 10^7$~yr. Thus the emission from stars younger 
than $t_0$ is more attenuated than that from older stars. Following 
\citet{CF00}, we express the luminosity per unit wavelength emerging 
at time $t$ from a galaxy as
\begin{equation}
L_{\lambda}(t)= \int_{0}^{t} dt' \psi(t-t')\,S_{\lambda}(t')\,
e^{-\hat{\tau}_{\lambda}(t')}\,,
\label{llambda}
\end{equation}
where $\psi(t-t')$ is the star formation rate at time $t-t'$,
$S_{\lambda}(t')$ is the luminosity per unit wavelength per unit mass 
emitted by a stellar generation of age $t'$, and $\hat{\tau}_{\lambda}
(t')$ is the `effective' absorption optical depth of the dust seen by 
stars of age $t'$ (i.e. averaged over photons emitted in all directions
by stars in all locations within the galaxy). The time dependence of
$\hat{\tau}_{\lambda}$ reflects the different attenuation affecting 
young and old stars in galaxies, 
\begin{eqnarray}
\hat{\tau}_\lambda(t')=\cases{
\hat{\tau}_\lambda^{\,\mathrm{BC}}+\hat{\tau}_\lambda^{\,\mathrm{ISM}}\,
&for $t'\leq t_0$,\cr
\hat{\tau}_\lambda^{\,\mathrm{ISM}}\,
&for $t'> t_0$.\cr
}
\label{taueff}
\end{eqnarray}
Here $\hat{\tau}_\lambda^{\,\mathrm{BC}}$ is the effective absorption 
optical depth of the dust in stellar birth clouds and $\hat{\tau}_\lambda^{\,
\mathrm{ISM}}$ that in the ambient ISM. We also adopt the prescription of 
\citet{CF00} to compute the emergent luminosities $L_\mathrm{H\alpha}(t)$ and
$L_\mathrm{H\beta}(t)$ of the H$\alpha$ ($\lambda=6563$ \AA) and H$\beta$ 
($\lambda=4861$ \AA) Balmer lines of hydrogen produced by stars in the birth
clouds. This assumes case~B recombination and includes the possible absorption
of ionising photons by dust before they ionise hydrogen.

The shape of the effective absorption curve depends on the combination 
of the optical properties and spatial distribution of the dust. We adopt 
the following dependence of $\hat{\tau}_\lambda^{\,\mathrm{BC}}$ and 
$\hat{\tau}_\lambda^{\,\mathrm{ISM}}$ on wavelength:
\begin{equation}
\hat{\tau}_\lambda^{\,\mathrm{BC}}=(1-\mu)\hat{\tau}_V\left(\lambda/{5500\,\mathrm{\AA}}\right)^{-1.3}\,,
\label{tau_bc}
\end{equation}
\begin{equation}
\hat{\tau}_\lambda^{\,\mathrm{ISM}}=\mu \hat{\tau}_V\left(\lambda/{5500\,\mathrm{\AA}}\right)^{-0.7}\,,
\label{tau_ism}
\end{equation}
where $\hat{\tau}_{V}$ is the total effective {\it V}-band absorption
optical depth of the dust seen by young stars inside birth clouds, and
$\mu=\hat{\tau }_V^{\,\mathrm{ISM}}/(\hat{\tau}_V^{\,\mathrm{BC}}+\hat{\tau
}_V^{\,\mathrm{ISM}})$ is the fraction of this contributed by
dust in the ambient ISM. The dependence of $\hat{\tau}_\lambda^{\,
\mathrm{ISM}}$ on $\lambda^{-0.7}$ is well constrained by the observed relation 
between ratio of far-infrared to ultraviolet luminosity and ultraviolet
spectral slope for nearby starburst galaxies (see \citealt{CF00}). The
dependence of $\hat{\tau}_\lambda^{\,\mathrm{BC}}$ on wavelength is
less constrained by these observations, because stellar birth clouds tend to
be optically thick, and hence, stars in these clouds contribute very little 
to the emergent radiation (except in the emission lines). For simplicity,
\cite{CF00} adopt $\hat{\tau}_\lambda^{\,\mathrm{BC}}\propto\lambda^{-0.7}$
by analogy with $\hat{\tau}_\lambda^{\,\mathrm{ISM}}$. We adopt here a 
slightly steeper dependence, $\hat{ \tau}_\lambda^{\, \mathrm{BC}} 
\propto \lambda^{-1.3}$ (equation \ref{tau_bc}), which corresponds to the
middle range of the optical properties of dust grains between the Milky Way,
the Large and the Small Magellanic Clouds (see section 4 of \citealt{CF00}).
This choice is motivated by the fact that giant molecular clouds can be
assimilated to foreground shells when attenuating the light from newly born stars.
In this case, the effective absorption curve should reflect the actual 
optical properties of dust grains. We emphasise that the dependence of
$\hat{\tau}_\lambda^{\, \mathrm{BC}}$ on wavelength has a negligible 
influence on the emergent ultraviolet and optical continuum radiation. It
affects mainly the attenuation of emission lines in galaxies with large
$\hat{\tau}_V^{\,\mathrm{BC}}/ \hat{ \tau}_V^{\,\mathrm{ISM}}$ and hence 
small $\mu$ (Section \ref{attenuation_law} below; see also \citealt{WILD07}). 

The fraction of stellar radiation absorbed by dust in the stellar birth clouds
and in the ambient ISM is reradiated in the infrared. We write the total
luminosity absorbed and reradiated by dust as the sum
\begin{equation}
\ldust(t)= \ldbc(t)+\ldism(t)\,,
\label{ldust}
\end{equation}
where 
\begin{equation}
\ldbc(t)=\int_{0}^{\infty} d\lambda \left(1-e^{-\hat{\tau}_{\lambda}^\mathrm{BC}}\right) \int_{0}^{t_0} dt' \psi(t-t') S_{\lambda}(t')
\label{ldust_bc}
\end{equation}
is the total infrared luminosity contributed by dust in the birth clouds, and 
\begin{equation}
\ldism(t)=\int_{0}^{\infty} d\lambda \left(1-e^{-\hat{\tau}_{\lambda}^\mathrm{ISM}}\right) \int_{t_0}^{t} dt' \psi(t-t') S_{\lambda}(t')
\label{ldust_ism}
\end{equation}
is the total infrared luminosity contributed by dust in the ambient ISM. For 
some purposes, it is also convenient to define the fraction of the total 
infrared luminosity contributed by the dust in the ambient ISM:
\begin{equation}
f_\mu(t) \equiv {\ldism(t)/\ldust}(t)\,.
\label{fmu1}
\end{equation}
This depends on the total effective {\it V}-band absorption optical depth of
the dust, \tauv, the fraction $\mu$ of this contributed by dust in the ambient
ISM, and the star formation history (and IMF) determining the relative 
proportion of young and old stars in the galaxy.


	\subsection{Infrared emission of the dust}
	\label{irmodel}

We now present a simple but physically motivated prescription to compute 
the spectral distribution of the energy reradiated by dust in the 
infrared (i.e., the distribution in wavelength of $L_{\mathrm{d}}^{\, 
\mathrm{BC}}$ and $L_{\mathrm{d} }^{\, \mathrm{ISM}}$). By construction,
the infrared emission computed in this way can be related to the emission
at shorter wavelengths using equations (\ref{llambda})--(\ref{ldust_ism})
above.

	\subsubsection{Components of infrared emission}
	\label{components}
	
\begin{figure*}
\centering
\includegraphics[width=0.75\textwidth]{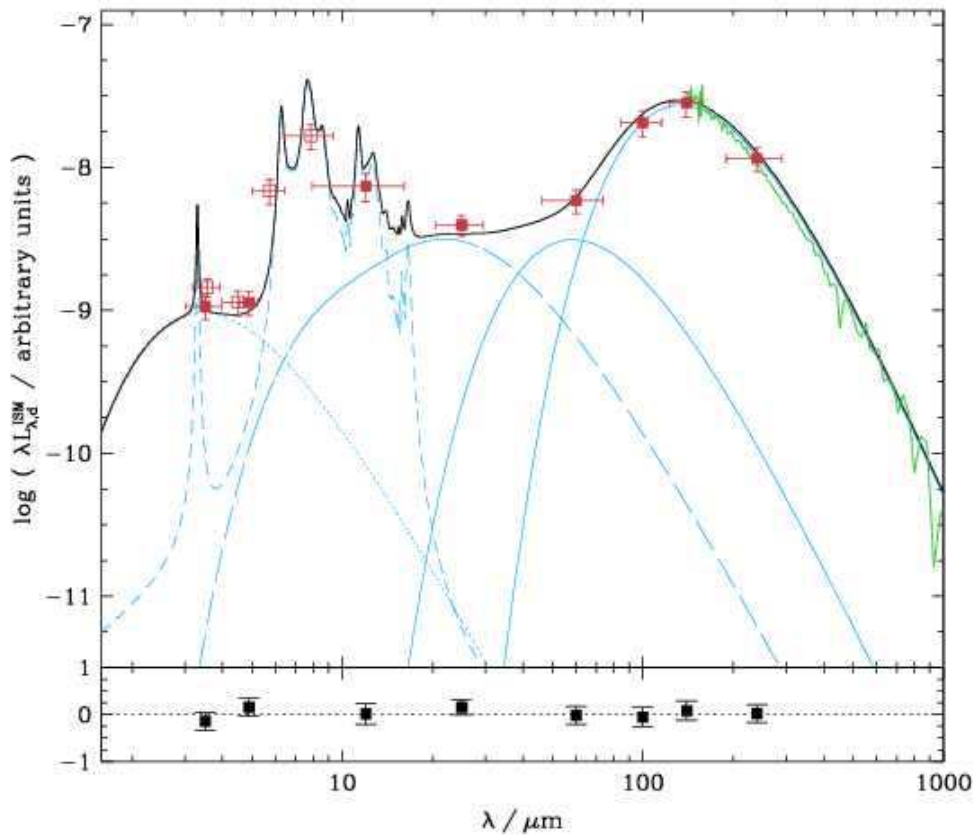}
\caption{Best model fit (in black) to the observed mean spectral energy
distribution of the Galactic cirrus emission. The model was computed using
equation~(\ref{ldust_ism2}) and the parameters listed in 
equation~(\ref{ism_parameters}). The red filled squares are the {\it
COBE}/DIRBE observations of \citet{DWEK97}. Also shown for reference are the
{\it Spitzer}/IRAC observations of \citet[red open squares]{FLAG06} and
the {\it COBE}/FIRAS observations of \citet[green line]{DWEK97}. The blue
lines show the decomposition of the model in its different components (see
Section~\ref{irmodel}): near-infrared continuum ({\it dotted}); PAHs ({\it
short-dashed}); hot mid-infrared continuum ({\it long-dashed}) and warm and
cold grains in thermal equilibrium ({\it solid}). The fit residuals 
$(L_\lambda^\mathrm{obs} - L_\lambda^\mathrm{mod})/ L_\lambda^\mathrm{obs}$,
where $L_\lambda^\mathrm{obs}$ and $L_\lambda^\mathrm{mod}$ are the observed
and model fluxes in a given photometric band, are shown at the bottom.}
\label{fit_ism}
\end{figure*}

The infrared emission from galaxies is generally attributed to three main
constituents of interstellar dust: polycyclic aromatic hydrocarbons (PAHs),
which produce strong emission features at wavelengths between 3
and 20~\mic; `very small grains' (with sizes typically less than 0.01~\mic),
which are stochastically heated to high temperatures by the absorption
of single ultraviolet photons; and `big grains' (with sizes typically between
0.01 and 0.25~\mic), which are in thermal equilibrium with the radiation field.
This picture arises from detailed models of the sizes and optical properties
of dust grains and PAH molecules in the ISM of the Milky Way and other nearby
galaxies (e.g., \citealt{MRN77,DL84,LP84}). Here, we build on the results from
these studies to describe the different components of infrared emission in
galaxies, without modelling in detail the physical properties of dust grains. 
As mentioned in Section \ref{intro}, the motivation for this approach is to
build a model simple enough to derive statistical constraints on the star 
formation and dust properties of large samples of observed galaxies, based on 
consistent fits of the ultraviolet, optical and infrared emission. We describe 
the different components of infrared emission in our model as follows:

	\begin{enumerate}

\item {\it PAH and near-infrared continuum emission.}
The mid-infrared spectra of most normal star-forming 
galaxies are dominated by strong emission features at 3.3, 6.2, 7.7, 8.6, 
11.3 and 12.7~\mic. Although still uncertain, the carriers of these features
are generally accepted to be PAH molecules transiently excited to high 
internal energy levels after the absorption of single ultraviolet photons
\citep{LP84,ALL85,LHD89,ALL99}. PAH emission tends to peak in the 
`photo-dissociation regions' (PDRs) at the interface between ionised 
and molecular gas in the outskirts of \hii \ regions, where PAH molecules
can survive and transient heating is most efficient (e.g., \citealt{C96,
VER96,RJB05}). In these environments, the non-ionising ultraviolet radiation
from young stars dominates the energy balance and can dissociate molecules
such as H$_2$ and CO (see \citealt{PDR} for a review).

Observations with {\it ISO}/ISOPHOT of the mid-infrared spectra of
45 nearby, normal star-forming galaxies do not reveal any strong
variation of the PAH emission spectrum with galaxy properties such as 
infrared colours and infrared-to-blue luminosity ratio \citep{HEL00}.
For simplicity, we adopt a fixed template spectrum to describe PAH 
emission in our model.\footnote{In reality, changes in PAH molecule 
size and ionisation state, variations in metallicity and contamination
by an AGN could cause variations of up to 40 per cent in the relative 
strengths of some PAH emission features \citep{HEL00,B06,S07, GAL07}. We 
ignore this refinement here.} We use the mid-infrared spectrum of the 
photo-dissociation region in the prototypical Galactic star-forming region
M17 SW extracted from the {\it ISO}/ISOCAM observations of \cite{C96} by
\cite{MADD06}. The observed spectrum does not extend blueward of 5~\mic.
We extend it to include the 3.3~\mic\ PAH emission feature using the 
Lorentzian profile parametrised by \cite{VER01}, $F^{\mathrm{Lorentz}}_\nu
=f_0 [1+(x-x_0)^2/ \sigma^2]^{-1}$, where $x=1/\lambda$ is the wavenumber,
$x_0=3039.1$~cm$^{-1}$ the central wavenumber of the feature and $\sigma=
19.4$~cm$^{-1}$ half the FWHM. We set the amplitude $f_0$ so that the 
luminosity of the  3.3~\mic\ emission feature is 10 per cent of that of the
11.3~\mic\ feature (the relative ratio found by \citealt{LD01} for neutral
PAHs). We write the spectral energy distribution of PAHs in our model
\begin{equation}
l_{\lambda}^{\mathrm{PAH}}={L_{\lambda}^{\mathrm{M17}}}
\left(
\int_0^\infty d\lambda\,{L_{\lambda}^{\mathrm{M17}}}
\right)^{-1} \,,
\label{lpah}
\end{equation}
where $L_{\lambda}^{\mathrm{M17}}$ is the adopted M17 spectral template,
and $l_{\lambda}^{\mathrm{PAH}}$ is normalised to unit total energy.

The above {\it ISO}/ISOPHOT observations of a sample of 45 normal
star-forming galaxies also reveal a component of near-infrared continuum 
emission characterised by a very high colour temperature ($\sim1000$~K)
at wavelengths between 3 and 5~\mic\ \citep{LU03}. This component accounts
typically for at most a few percent of the total infrared luminosity,
but it contributes significantly to the observed {\it Spitzer}/IRAC band
fluxes at 3.6 and 4.5~\mic. It is also present in the spectra of reflection
nebulae \citep{SELL83} and in the diffuse cirrus emission of the Milky Way
\citep{DWEK97,FLAG06}. The origin of this emission is still uncertain, and 
it could be related to the stochastic heating of PAH molecules or carbon 
grains (e.g.  \citealt{FLAG06}). \cite{LU03} find that the strength of this
emission correlates well with that of PAH features in the spectra of the 
galaxies in their sample.  In particular, the ratio of the continuum flux
density at 4~\mic\ to the mean flux density of the 7.7~\mic \ PAH feature
has a roughly constant value of 0.11. 

To implement this component of near-infrared continuum emission associated
to PAH emission in our model, we use a greybody (i.e. modified blackbody) 
function of the form (e.g., \citealt{HIL83}) 
\begin{equation}
l_\lambda^{T_{\mathrm{d}}}=
\kappa_\lambda\,B_{\lambda}(T_{\mathrm{d}})
\left[
\int_0^\infty d\lambda\,\kappa_\lambda\,B_{\lambda}(T_{\mathrm{d}})
\right] ^{-1} \,,
\label{mbb}
\end{equation}
where $B_{\lambda}(T_{\mathrm{d}})$ is the Planck function of temperature
$T_{\mathrm{d}}$, $\kappa_\lambda$ is the dust mass absorption coefficient,
and $l_\lambda^{T_{\mathrm{d}}}$ is normalised to unit total energy. The 
dust mass absorption coefficient is usually approximated by a single power
law of the form
\begin{equation}
\kappa_\lambda\propto\lambda^{-\beta}\,,
\label{kappa}
\end{equation}
where $\beta$ is the dust emissivity index. Models and observations
of infrared to sub-millimetre dust emission and laboratory studies of
carbonaceous and amorphous silicate grains suggest values of $\beta$
in the range $1\la \beta\la2$, with some dependence on grain size and
temperature (\citealt{AND74,HIL83,DL84,REACH95,AG96, BOUL96,M98}; see
also section 3.4 of \citealt{DUNNE01} and section 4 of \citealt{DH02}
for discussions on the constraints on $\beta$).
Typically, these studies favour $\beta \approx1$ for small carbonaceous
grains, which radiate most of their energy at mid-infrared wavelengths,
and $\beta\approx1.5 - 2$ for big silicate grains, which reach lower
temperatures and radiate most of their energy at far-infrared and 
sub-millimeter wavelengths. We therefore adopt $\beta=1$ in 
equation~(\ref{kappa}) to compute the near-infrared continuum emission 
in our model. We scale this emission so that the continuum flux
density at 4~\mic\ be 0.11 times the mean flux density of the 7.7~\mic \
PAH feature (see above). We find that a temperature $T_{\mathrm{d}}=850$~K
provides optimal fits to the observed spectral energy distributions of
galaxies at wavelengths between 3 and 5~\mic\ (see, e.g., Fig.~\ref{fit_ism}).

\item {\it Mid-infrared continuum emission from hot dust.}
In addition to PAH features, the mid-infrared spectra of star-forming 
galaxies (out to wavelengths $\lambda\sim40\,\mic$) also include a 
component of smooth continuum emission. This component is generally 
attributed to a continuous distribution of small grains with very low
heat capacity, which are stochastically heated to high temperatures by 
the absorption of single ultraviolet photons (e.g., \citealt{SEL84}).
The accurate modelling of this emission would require the computation
of temperature fluctuations of grains of different sizes and optical 
properties caused by the absorption of ultraviolet photons in 
different interstellar radiation fields (e.g.  \citealt{P76,AK79}). For
simplicity, in our model, we describe the `hot' mid-infrared continuum 
emission as the sum of two greybodies (equation~\ref{mbb}) of temperatures
$T_{\mathrm{d}}=130$ and 250~K, with equal contributions to the total
infrared luminosity,
\begin{equation}
l_\lambda^\mathrm{\,MIR}=
(l_\lambda^{250}+l_\lambda^{130})
\left[
\int_0^\infty d\lambda\,(l_\lambda^{250}+l_\lambda^{130})
\right] ^{-1} \,.
\label{lmir}
\end{equation}
We find that these two temperatures reproduce in an optimal way the range of 
infrared colours of star-forming galaxies (see Section~\ref{colours} below).

\item {\it Emission from grains in thermal equilibrium.} 
At far-infrared wavelengths, the emission from galaxies is generally 
dominated by dust grains in thermal equilibrium at low temperatures. The grain
temperatures depend sensitively on the intensity of the interstellar radiation
field. This is why the peak of the far-infrared spectral energy distribution
of a galaxy is a good indicator of dust heating in the ISM.

We consider two types of grains in thermal equilibrium in our model:
warm grains, which can reside both in stellar birth clouds and in the
ambient ISM, with characteristic temperatures \tbgswarm\ and \tbgswarmism,
respectively; and cold grains, which can reside only in the ambient ISM, with
characteristic temperature \tbgscold.  To account for the observed dispersion
in the peak of the far-infrared emission of galaxies with different
star formation activities, we allow the warm-grain temperature to vary between
30 and 60~K and the cold-grain temperature to vary between 15 and 25~K. We
describe the emission from grains in thermal equilibrium using greybody 
spectra (equation~\ref{mbb}) with emissivity index $\beta=1.5$ for warm dust
and $\beta=2.0$ for cold dust (equation \ref{kappa}).

\end{enumerate}

In summary, the infrared spectral energy distribution of 
stellar birth clouds in our model can be written
\begin{equation}
L_{\lambda,\mathrm{d}}^{\mathrm{\,BC}}=\big(\xipahs\,l_{\lambda}^{\mathrm{PAH}}+
\ximir\,l_{\lambda}^\mathrm{\,MIR}+
\xiwarm\,l_{\lambda}^\mathrm{\tiny \tbgswarm}\big)\, 
\big(1-f_\mu\big)\ldust \,,
\label{ldust_bc2}
\end{equation}
where \ldust \ is the total infrared luminosity reradiated by dust
(equation \ref{ldust}), \fmu \ is the fraction of this contributed 
by the ambient ISM (equation \ref{fmu1}), $l_{\lambda}^{\mathrm{PAH}}$,
$l_{\lambda}^{\mathrm{\,MIR}}$ and $l_{\lambda}^\mathrm{\tbgswarm}$
are computed using equations (\ref{lpah}), (\ref{lmir}) and (\ref{mbb}), 
and \xipahs, \ximir\ and \xiwarm\ are the relative contributions by
PAHs, the hot mid-infrared continuum and grains in thermal equilibrium 
to the total infrared luminosity of the birth clouds. These satisfy 
the condition
\begin{equation}
\xipahs+\ximir+\xiwarm=1\,.
\label{xisumbc}
\end{equation}

Similarly, the infrared spectral energy distribution of the ambient
ISM can be written
\begin{eqnarray}
\begin{array}{rcl}
L_{\lambda,\mathrm{d}}^{\mathrm{\,ISM}} &=& 
\big(
\xipahsism\,l_{\lambda}^{\mathrm{\,PAH}}+
\ximirism\,l_{\lambda}^{\mathrm{\,MIR}}+
\xiwarmism\,l_\lambda^{\tiny \tbgswarmism}
{} \\ & & {}
+ \xicold\,l_\lambda^{\tiny \tbgscold}
\big)\, f_\mu \ldust \,,
\label{ldust_ism2}
\end{array}
\end{eqnarray}
where $l_{\lambda}^{\mathrm{\,PAH}}$, $l_{\lambda}^{\mathrm{\,MIR}}$,
$l_{\lambda}^{\tiny \tbgswarmism}$ and $l_{\lambda}^{\tiny \tbgscold}$
are computed using equations~(\ref{lpah}), (\ref{lmir}) and (\ref{mbb}),
and \xipahsism, \ximirism, \xiwarmism\ and \xicold\ are the relative 
contributions by PAHs, the hot mid-infrared continuum and warm and cold 
grains in thermal equilibrium to the total infrared luminosity of the 
ISM. These satisfy the condition
\begin{equation}
\xipahsism+\ximirism+\xiwarmism+\xicold=1\,.
\label{xisumism}
\end{equation}

In practice, we can fix the shape of the mid-infrared spectral energy 
distribution of the ambient ISM in our model to keep the number of 
adjustable parameters as small as possible. This is justified by the
fact that the intensity of the average radiation field heating dust 
in the diffuse ISM of normal galaxies is roughly constant (e.g., 
\citealt{H86}). Moreover, sophisticated models of dust emission by 
\cite{LD01} and \cite{DL07} suggest that even large variations of 
the intensity of the interstellar radiation field have only a
small influence on the overall shape of the diffuse mid-infrared 
spectral energy distribution. In these conditions, we can appeal for
example to observations of high-Galactic-latitude (cirrus) dust 
emission in the Milky Way to constrain the mid-infrared spectral 
energy distribution of the ambient ISM in our model. In 
Fig.~\ref{fit_ism}, we show the model spectral energy distribution
$L_{\lambda,\mathrm{d}}^{\mathrm{\,ISM}}$ (in black)
computed using equation~(\ref{ldust_ism2}) that best fits the mean
Galactic cirrus emission observed by the {\it Cosmic Background 
Explorer}/Diffuse Infrared Background Experiment ({\it COBE}/DIRBE)
at wavelengths between 3.5 and 240~\mic\ \citep{DWEK97}. 
The blue lines show the different components of the best-fit model, with
parameters
\begin{eqnarray}
\xipahsism&=&0.22\cr
\ximirism&=&0.11\cr
\xiwarmism&=&0.07\cr
\xicold&=&0.60\cr
\tbgswarmism&=&45\mathrm{\,K}\cr
\tbgscold&=&18\mathrm{\,K}\,.
\label{ism_parameters}
\end{eqnarray}
For reference, we include in Fig.~\ref{fit_ism} additional observations 
of the mean Galactic cirrus emission obtained with {\it Spitzer}/IRAC 
at wavelengths 3.6, 4.5, 5.8 and 8.0~\mic\ \citep{FLAG06} and with the
{\it COBE}/Far-Infrared Absolute Spectrophotometer (FIRAS) at wavelengths
between 140 and 1000~\mic\ \citep{DWEK97}. The model reproduces these
observations remarkably well, even though they were not included in 
the fit. 

We use the constraints of Fig.~\ref{fit_ism} to fix the mid-infrared 
spectral energy distribution of the ambient ISM and reduce the number
of adjustable parameters in our model. Specifically, we fix the relative
contributions by PAHs, the hot mid-infrared continuum and warm grains in
thermal equilibrium to the total infrared luminosity of 
the ambient ISM (i.e. the relative ratios of \xipahsism, \ximirism\ and
\xiwarmism) to their values of equation~(\ref{ism_parameters}), and we
keep \xicold\ and \tbgscold\ as adjustable parameters. Thus, for a given 
contribution \xicold\ by cold grains in thermal equilibrium to the total
infrared luminosity of the ambient ISM, the contributions by PAHs, the
hot mid-infrared continuum and warm grains in thermal equilibrium are 
$\xipahsism=0.550(1-\xicold)$, $\ximirism=0.275 (1-\xicold)$ and 
$\xiwarmism=0.175(1-\xicold)$, respectively. 

The total spectral energy distribution of a galaxy in our model is 
computed as the sum
\begin{equation}
L_{\lambda,\mathrm{d}}^{\mathrm{\,tot}} = 
L_{\lambda,\mathrm{d}}^{\mathrm{\,BC}} +
L_{\lambda,\mathrm{d}}^{\mathrm{\,ISM}}\,,
\label{irsed}
\end{equation}
where $L_{\lambda,\mathrm{d}}^{\mathrm{\,BC}}$ and 
$L_{\lambda,\mathrm{d}}^{\mathrm{\,ISM}}$ are given by 
equations~(\ref{ldust_bc2}) and (\ref{ldust_ism2}). For some purposes, 
it is also convenient to define the global contribution by a specific
dust component, including stellar birth clouds and the ambient ISM,
to the total infrared luminosity of a galaxy. This can be written
\begin{equation}
\xipahstot =\xipahs\,(1-\fmu)+0.550\,(1-\xicold)\,\fmu\,,
\label{exipahstot}
\end{equation}
\begin{equation}
\ximirtot=\ximir\,(1-\fmu)+0.275\,(1-\xicold)\,\fmu\,,
\label{eximirtot}
\end{equation}
\begin{equation}
\xiwarmtot=\xiwarm\,(1-\fmu)+0.175\,(1-\xicold)\,\fmu\,,
\label{exiwarmtot}
\end{equation}
\begin{equation}
\xi_\mathrm{C}^\mathrm{\,tot}=\xicold\,\fmu\,,
\label{exicoldtot}
\end{equation}
for PAHs, the hot mid-infrared continuum and warm and cold dust in thermal
equilibrium, respectively.

We compute luminosity densities in any filter from $L_{\lambda, 
\mathrm{d}}^{\mathrm{\,tot}}$ using the general formula
\begin{equation}
\displaystyle
L_\nu^{\lambda_0} = C_{\nu_0}
\frac{\int d\nu\,\nu^{-1}\,L_\nu\,R_\nu}
     {\int d\nu\,\nu^{-1}\,C_\nu\,R_\nu}
= \frac{\lambda_0^2}{c}\,C_{\lambda_0} \frac{\int
d\lambda\,L_\lambda\,\lambda\,R_\lambda}{\int
d\lambda\,C_\lambda\,\lambda\,R_\lambda}\,,
\label{flux-density}
\end{equation}
where
\begin{equation}
\lambda_0=\frac{\int d\lambda\,\lambda\,R_\lambda}{\int
d\lambda\,R_\lambda}
\end{equation}
is the effective wavelength of the filter of response $R_\lambda$ ($R_\nu$),
and the calibration spectrum $C_\lambda$ ($C_\nu$) depends on the photometric
system. For the AB system, $C_\nu$ is fixed at 3631~Jy, implying $C_\lambda
\propto \lambda^{-2}$ \citep{OG83}. For the {\it IRAS}, {\it ISO}/ISOCAM and
{\it Spitzer}/IRAC photometric systems, the convention is to use $\nu 
C_\nu=\lambda C_\lambda=\,$constant, and hence, $C_\lambda \propto 
\lambda^{-1}$ (see \citealt{IRAS,ISOCAM,IRAC_PHOT}, and also the IRAC Data
Handbook\footnote{http://ssc.spitzer.caltech.edu/irac/dh/}). The 
{\it Spitzer}/MIPS system was calibrated using a blackbody spectrum of 
temperature 10,000~K, such that $C_\lambda=B_\lambda(10,000\,{\mathrm K)}$
(MIPS Data Handbook\footnote{http://ssc.spitzer.caltech.edu/mips/dh/}).
\begin{figure*}
\centering
\includegraphics[width=0.75\textwidth]{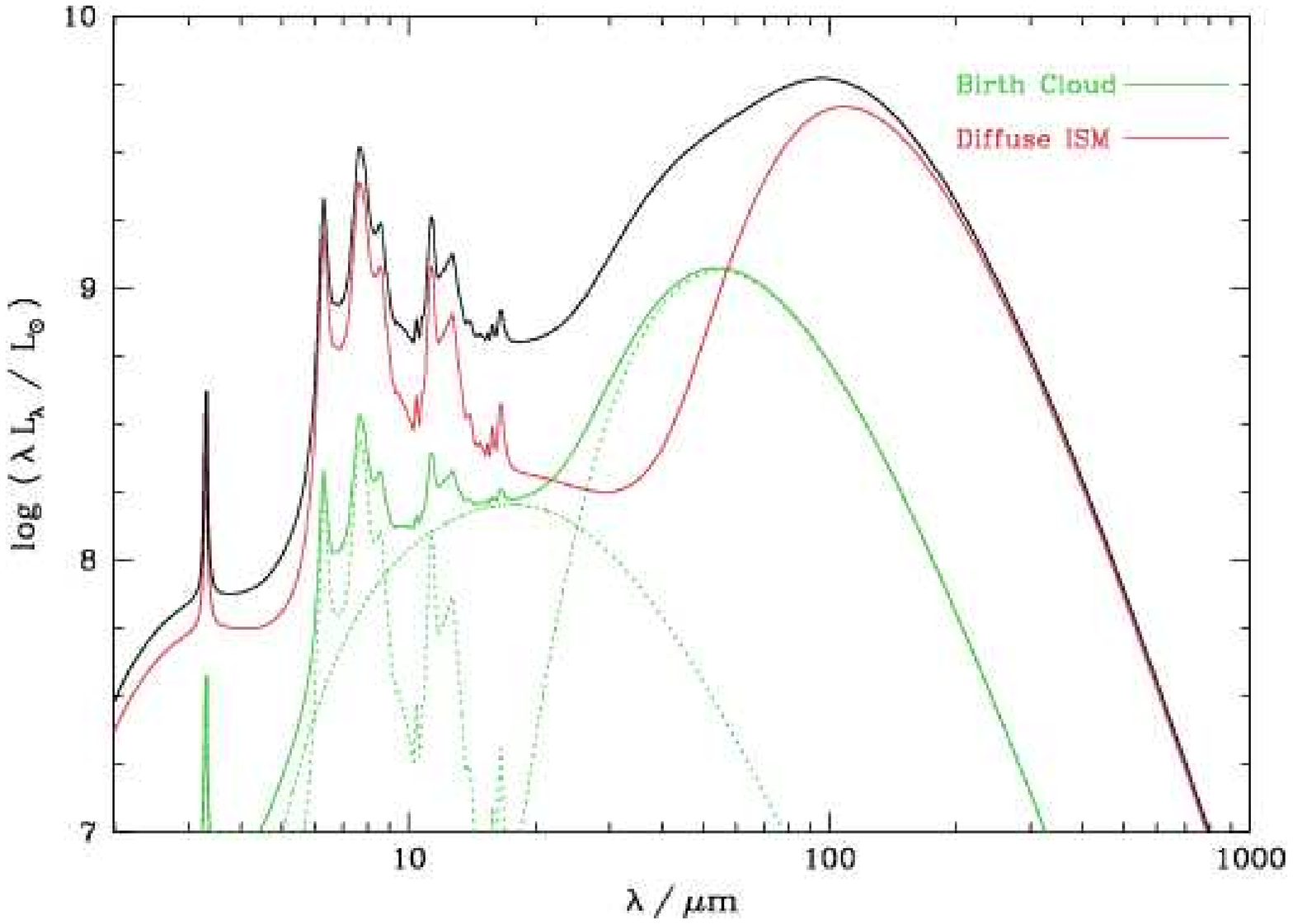}
\caption{Example of infrared spectral energy distribution computed 
using the model presented in Section \ref{components}, for `standard'
values of the parameters (equation \ref{standard}): $f_\mu=0.6$, $\xipahs
=0.05$, $\ximir=0.15$, $\xiwarm=0.80$, $\tbgswarm=48\mathrm{\,K}$, 
$\xicold=0.8$ and $\tbgscold=22\mathrm{\,K}$. The black curve shows the
total infrared spectrum. The green solid curve shows the contribution by
dust in the stellar birth clouds, and the green dashed curves the
breakdown of this contribution in different components: PAHs, mid-infrared
continuum and warm dust in thermal equilibrium. The red curve shows the
contribution by dust in the ambient ISM. The total luminosity is 
$\ldust=10^{10} L_{\sun}$ (equations \ref{ldust_bc2}, \ref{ldust_ism2} 
and \ref{irsed}).}
\label{eg_ir_sed}
\end{figure*}
In Fig.~\ref{eg_ir_sed}, we show an example of infrared spectral energy
distribution $L_{\lambda,\mathrm{d}}^{\mathrm{\,tot}}$ computed using our
model, for parameter values typical of normal star-forming galaxies 
(equation~\ref{standard} below). As we shall see in Section~\ref{colours},
this parameterisation of infrared galaxy spectra allows us to account for
the full range of observed colours of star-forming galaxies.

\subsubsection{Comparison with observed infrared colours}
\label{colours}

To test how well our model can reproduce the observed infrared colours
of galaxies in a wide range of star formation histories, we appeal to 
a sample of 157 nearby galaxies compiled by \cite{DE02}, for which {\it 
IRAS} and {\it ISO}/ISOCAM observations are available from \cite{BOSELLI98}, 
\cite{ROUSSEL01}, \cite{LAURENT00} and \cite{DALE00}. The galaxies in 
this sample span wide ranges of morphologies, absolute infrared 
luminosities, infrared-to-blue luminosity ratios and infrared colours
(see \citealt{DE02} for more detail). For nearby galaxies, the flux 
density collected by the ISOCAM {\it LW2} (6.75~\mic) filter, \fseven, 
tends to be dominated by the PAH emission features at 6.2, 7.7 and 
8.6~\mic. The {\it IRAS} 12-\mic\ flux density \ftwelve\ captures the PAH 
emission at 11.3 and 12.7~\mic\ and the mid-infrared continuum 
emission from hot dust. The ISOCAM {\it LW3} (15~\mic) and the {\it IRAS}
25-\mic\ flux densities, \ffifteen\ and \ftwofive, reflect primarily
the mid-infrared continuum emission from hot dust. At longer wavelengths, 
the {\it IRAS} 60-\mic\ flux density \fsixty\ samples the emission from
warm grains in thermal equilibrium in star-forming clouds, and the {\it 
IRAS} 100-\mic\ flux density \fhundred\ that from colder grains in 
thermal equilibrium in the ambient interstellar medium. To test our model,
we require photometric observations in all these filters. This reduces the
sample to 107 galaxies. For the purpose of comparisons between our 
(angle-averaged) model and observations, we neglect possible anisotropies
and equate flux ratios at all wavelengths to the corresponding luminosity
ratios. 

\begin{figure*}
\centering
\includegraphics[width=0.7\textwidth]{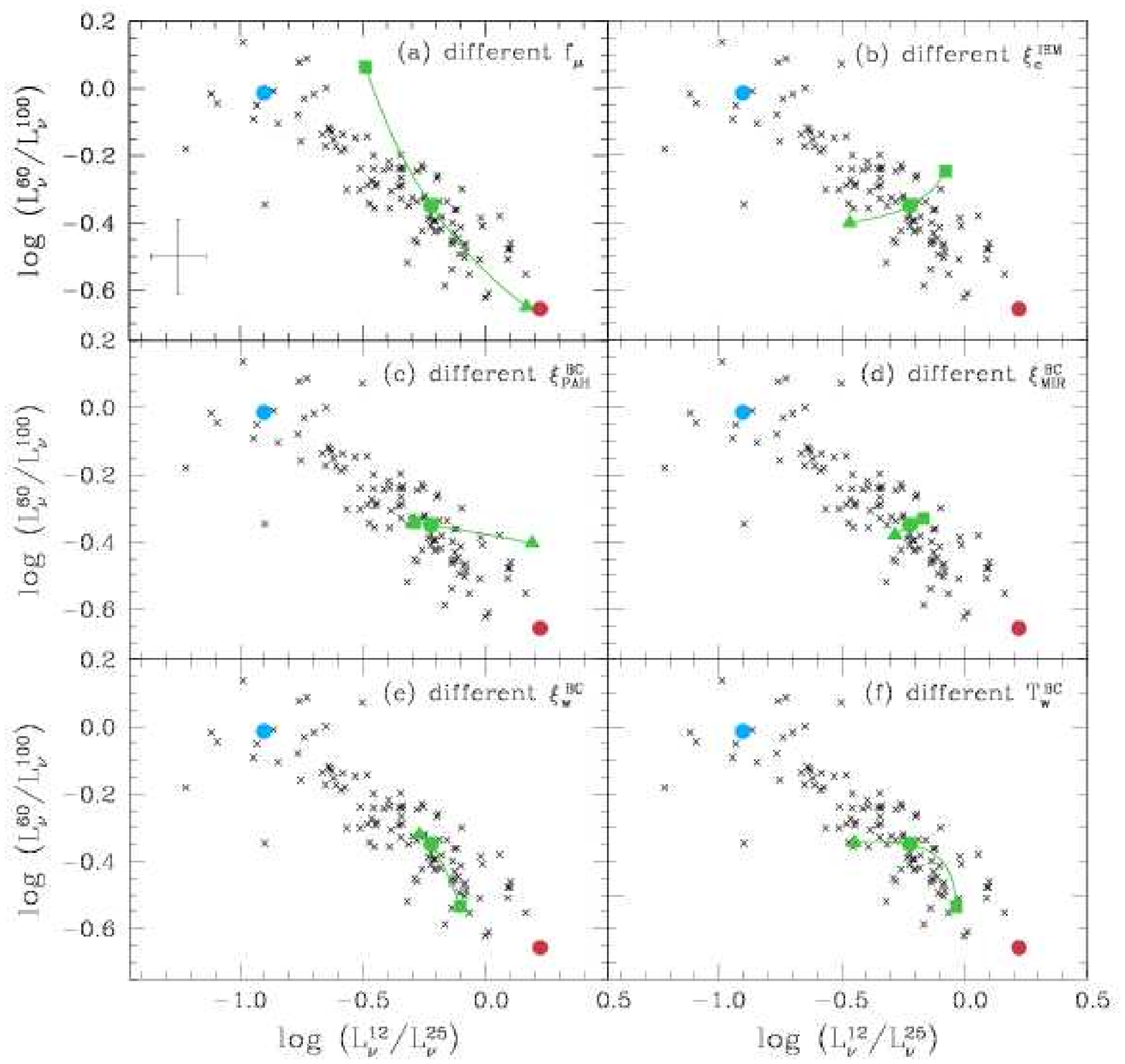}
\caption{Ratio of 60-\mic\ to 100-\mic\ {\it IRAS} luminosity density
plotted against ratio of 12-\mic\ to 25-\mic\ {\it IRAS} luminosity
density. The data points in all panels (black crosses) are from the
\protect\cite{DE02} sample discussed in Section~\ref{irmodel}
(typical measurement errors are indicated in the upper left panel).
In each panel, the green line shows the effect of varying one 
parameter of the model from the lower end of the range (square) to 
the standard value (green circle) to the upper end of the range 
(triangle), with all other parameters fixed at their standard values:
(a) fraction of total infrared luminosity contributed
by dust in the ambient ISM, \fmu\,= 0.05, 0.50 and 0.95; (b)
contribution by cold dust in thermal equilibrium to the infrared
luminosity of the ambient ISM, \xicold = 0.50, 0.80, 1.0;
(c) contribution by PAHs to the infrared luminosity of stellar birth 
clouds, \xipahs\ = 0.00, 0.05 and 0.50; (d) contribution by the hot
mid-infrared continuum to the infrared luminosity of stellar birth
clouds, \ximir  = 0.00, 0.15 and 0.50; (e) contribution by warm dust
in thermal equilibrium to the infrared luminosity of stellar
birth clouds, \xiwarm\ = 0.15, 0.80 and 0.95; (f) equilibrium
temperature of warm dust in stellar birth clouds, \tbgswarm\ = 
30, 48 and 60~K. In all panels, the red circle corresponds to the same
quiescent model galaxy dominated by cold dust emission, while the blue
circle corresponds to an actively star-forming galaxy dominated hot dust 
emission (see Table \ref{table_par} and Section~\ref{irmodel} for a 
description of these models).}
\label{varypar1}
\end{figure*}

\begin{figure*}
\centering
\includegraphics[width=0.7\textwidth]{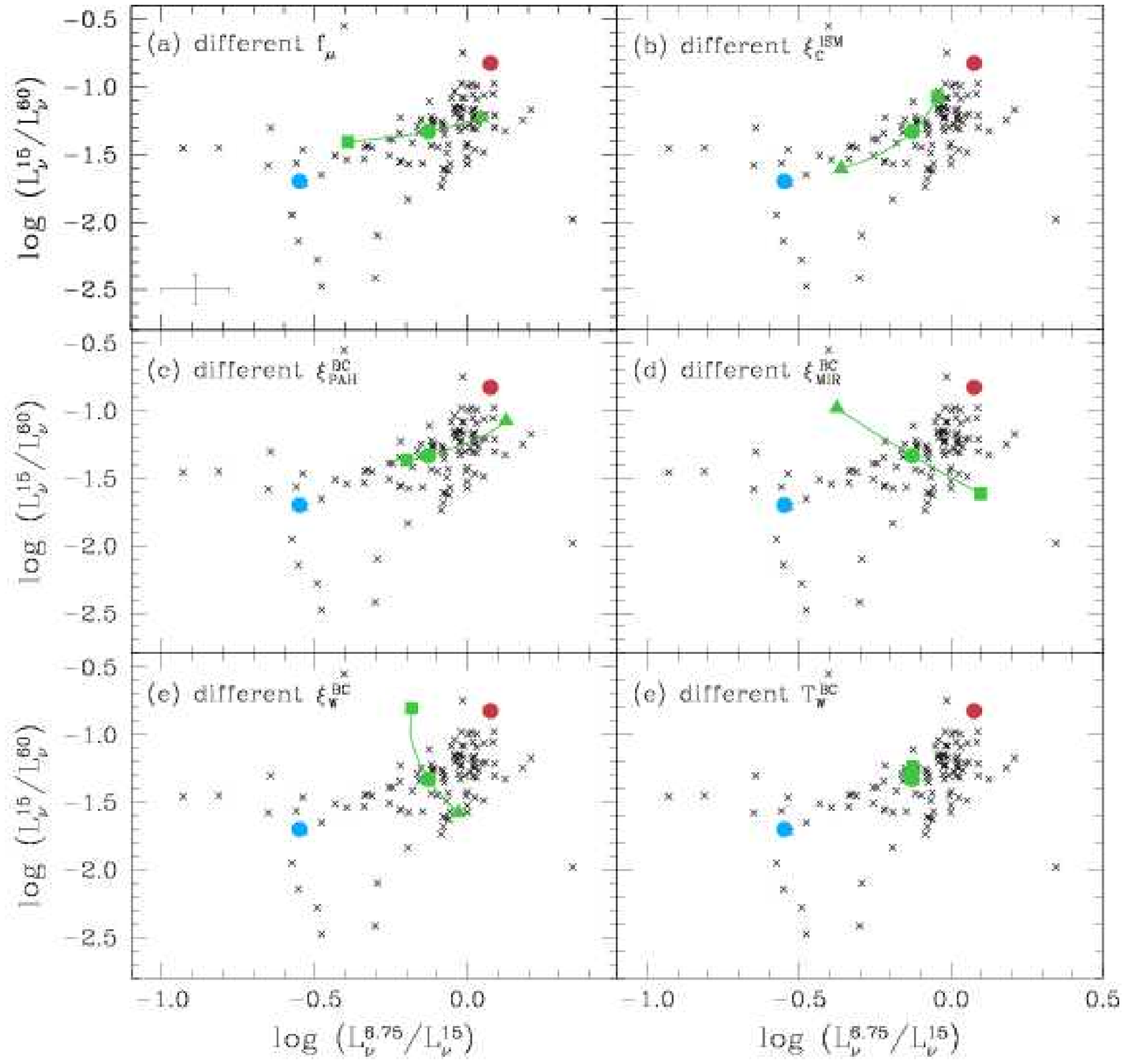}
\caption{Ratio of 15-\mic\ to 60-\mic\ {\it ISO} and {\it IRAS} luminosity
density plotted against ratio of 6.75-\mic\ to 15-\mic\ {\it ISO} 
luminosity density. The data points in all panels (black crosses) are 
from the \protect\cite{DE02} sample discussed in Section~\ref{irmodel}. 
In each panel, the green line shows the effect of varying one parameter 
of the model from the lower end of the range (square) to the standard 
value (green circle) to the upper end of the range (triangle), with all 
other parameters fixed at their standard values. The models are the same 
as in Fig.~\ref{varypar1}. In all panels, the red circle corresponds to 
the same quiescent model galaxy dominated by cold dust emission, while 
the blue circle corresponds to an actively star-forming galaxy dominated 
hot dust emission (see Table \ref{table_par} and Section~\ref{irmodel} for
a description of these models).
}
\label{varypar2}
\end{figure*}

In Figs.~\ref{varypar1} and \ref{varypar2}, we show the locations of
these galaxies in various infrared colour-colour diagrams (black crosses). 
Fig.~\ref{varypar1} shows $\lsixty/\lhundred$ as a function of $\ltwelve/
\ltwofive$ ({\it IRAS} colours), while Fig.~\ref{varypar2} shows 
$\lfifteen/\lsixty$ as a function of $\lseven/\lfifteen$ ({\it ISO} 
colours). The well-known correlations between the different infrared
colours of galaxies illustrated by Figs.~\ref{varypar1} and \ref{varypar2}
suggest that the different contributors to the total infrared emission
are related to one another. These relations are generally interpreted as
sequences in the overall star formation activity and dust heating (e.g.
\citealt{H86,DALE01}). Quiescent star-forming galaxies with strong PAH
emission and cool ambient-ISM dust tend to have high $\ltwelve/\ltwofive$,
$\lseven/\lfifteen$ and $\lfifteen/\lsixty$ and low $\lsixty/\lhundred$.
In contrast, actively star-forming galaxies, in which the mid-infrared
emission is dominated by continuum radiation by hot dust and the 
far-infrared emission by warm dust in star-forming regions, tend to 
have low $\ltwelve/\ltwofive$, $\lseven/\lfifteen$ and $\lfifteen/\lsixty$
and high $\lsixty/\lhundred$.

We now use these observations to explore the influence of each parameter
of our model on the various infrared colours. We explore the effect of
varying a single parameter at a time, keeping all the other parameters
fixed at `standard' values. After some experimentation, we adopted the
following standard parameters (corresponding to the model shown in 
Fig.~\ref{eg_ir_sed}):
\begin{eqnarray}
\fmu&=&0.60\cr
\xipahs&=&0.05\cr
\ximir&=&0.15\cr
\xiwarm&=&0.80\cr
\xicold&=&0.80\cr
\tbgswarm&=&48\mathrm{\,K}\cr
\tbgscold&=&22\mathrm{\,K}\,.
\label{standard}
\end{eqnarray}
These values allow the standard model to match roughly the observed 
typical (i.e. median) infrared colours of nearby star-forming galaxies
in Figs.~\ref{varypar1} and \ref{varypar2} (green circle). Each panel
of Figs.~\ref{varypar1} and \ref{varypar2} shows the effect of varying
one parameter of the model, with the other parameters held fixed at their
standard values.\footnote{In practice, an increase in any of \xipahs,
\ximir\ and \xiwarm\ will be accompanied by a drop in the other two 
fractions by virtue of equation~(\ref{xisumbc}). Likewise, an increase
in \xicold\ implies a drop in the sum $\xipahsism + \ximirism + 
\xiwarmism$ (equation~\ref{xisumism}). When exploring such variations, 
we keep the relative ratios of the unexplored fractions 
fixed.\label{xifracs}} We can summarise the role of each parameter as 
follows.

\begin{description}

\item{{\it Fraction of total infrared luminosity contributed by dust in the
ambient ISM.}}
The dominant effect of increasing \fmu\ is to increase the contribution to
the total infrared luminosity by cold dust. This raises the infrared 
luminosity around 100~\mic\ (Fig.~\ref{eg_ir_sed}), causing 
$\lsixty/\lhundred$ and even $\lfifteen/\lsixty$ to decrease 
(Figs.~\ref{varypar1}a and \ref{varypar2}a). In addition, a larger \fmu\
leads to an increase in $\ltwelve/\ltwofive$ and $\lseven/\lfifteen$, due
to the increased contribution by PAH emission features (which dominate 
the mid-infrared emission of the ISM) to the total mid-infrared emission.

\item{{\it Contribution by cold dust in thermal equilibrium to 
the infrared emission from the ambient ISM.}}
Increasing \xicold\ causes $\ltwelve/\ltwofive$ and $\lseven/\lfifteen$
to decrease in Figs.~\ref{varypar1}b and \ref{varypar2}b, because of the
corresponding drop in PAH emission from the ambient ISM (see 
footnote~\ref{xifracs}). Also, both $\lfifteen/\lsixty$ and 
$\lsixty/\lhundred$ decrease because of the larger contribution by 
cold dust to the 60~\mic\ flux, and even more so to the 100~\mic\ flux.

\item{{\it Contribution by PAHs to the infrared emission from stellar
birth clouds.}}
Increasing \xipahs\ raises the contribution by PAH features to \lseven\
and \ltwelve, and to a lesser extent \lfifteen\ (Fig.~\ref{eg_ir_sed}).
This leads to a marked increase in $\lseven/\lfifteen$ and $\ltwelve/
\ltwofive$ and a milder one in $\lfifteen/\lsixty$ in Figs.~\ref{varypar1}c
and \ref{varypar2}c. The slight decrease in $\lsixty/\lhundred$ when
\xipahs\ increases in Fig.~\ref{varypar1}c is caused by the associated drop
in \xibgsbc\ (footnote~\ref{xifracs}).

\item{{\it Contribution by the hot mid-infrared continuum to the infrared
emission from stellar birth clouds.}}
The main effect of increasing \ximir\ is to make \lfifteen\ and \ltwofive\
larger (Fig.~\ref{eg_ir_sed}). This causes a drop in $\ltwelve/\ltwofive$
and $\lseven/\lfifteen$, while $\lfifteen/\lsixty$ increases 
(Figs.~\ref{varypar1}d and \ref{varypar2}d). The slight decrease in 
$\lsixty/\lhundred$ when \ximir\ increases in Fig.~\ref{varypar1}c is again
caused by the associated drop in \xibgsbc\ (footnote~\ref{xifracs}). 

\item{{\it Contribution by warm dust in thermal equilibrium to the
infrared emission from stellar birth clouds.}}
Increasing \xiwarm\ makes \lsixty\ larger, and to a lesser extent also
\ltwofive\ and \lhundred\ (Fig.~\ref{eg_ir_sed}). Remarkably, this causes
$\ltwelve/\ltwofive$ to drop and $\lsixty/\lhundred$ to rise in 
Fig.~\ref{varypar1}e, almost along the observational relation. In 
Fig.~\ref{varypar2}e, an increase in \xiwarm\ makes $\lfifteen$ smaller
(because of the associated drop in \ximir; see footnote~\ref{xifracs}),
and hence, $\lfifteen/\lsixty$ smaller and $\lseven/ \lfifteen$ larger.

\item{{\it Equilibrium temperature of warm dust in stellar birth clouds.}}
Increasing \tbgswarm\ across the range from 30 to 60~K moves the wavelength
of peak luminosity of dust in thermal equilibrium in the stellar birth 
clouds roughly from 70 to 40~\mic. This causes $\ltwelve/\ltwofive$ to 
drop and $\lsixty/\lhundred$ to rise in Fig.~\ref{varypar1}f. In 
Fig.~\ref{varypar2}f, the effects on $\lseven/ \lfifteen$ and 
$\lfifteen/\lsixty$ are negligible. 

\item{{\it Equilibrium temperature of cold dust in the ambient ISM (not
shown).}}
The most significant effect of increasing \tbgscold \ at fixed \xicold\
in Figs.~\ref{varypar1} and \ref{varypar2} is a small rise in 
$\lsixty/\lhundred$ caused by a blue shift of the peak infrared luminosity
of cold dust in the ambient ISM (from about 140 to 80~\mic\ as \tbgscold \
increases from 15 to 25~K). Variations in \tbgscold\ also have a 
significant influence on the emission redward of 100~\mic.

\end{description}

The green lines in Figs.~\ref{varypar1} and \ref{varypar2} show that
the extremities of the observational relations between the different
infrared colours of galaxies cannot be reached by varying a single model 
parameter at a time. This suggests that variations in the different dust
components of galaxies are related to each other. However, we have 
checked that the properties of every galaxy in Figs.~\ref{varypar1} 
and \ref{varypar2} could be reproduced with at least one combination of
parameters of our model. We illustrate this by showing two models  
lying at the ends of the observational relations: a `cold' infrared 
spectrum characteristic of a quiescent galaxy with little star formation
(red circle); and a `hot' infrared spectrum characteristic of an 
actively star-forming, starburst galaxy (blue circle). The parameters
of these models are listed in Table~\ref{table_par}. We emphasise that
these are not unique sets of parameters optimised to fit specific 
galaxy spectral energy distributions. Rather, they are examples of how the
colours in those regions of the diagrams can be reproduced using our model.

Figs.~\ref{varypar1} and \ref{varypar2} allow us to draw some general
conclusions about the influence of the various parameters of our model
on the observed infrared colours of galaxies. For example, as expected
from Fig.~\ref{eg_ir_sed}, the $\lsixty/\lhundred$ colour appears to be
controlled primarily by the fraction \fmu\ of total infrared luminosity
contributed by dust in the ambient ISM and the properties of dust in 
thermal equilibrium: the relative contribution \xicold\ by cold dust
to the infrared luminosity of the ambient ISM (and the temperature 
\tbgscold\ of this dust ) and the relative contribution \xiwarm\ by 
warm dust to the infrared luminosity of stellar birth clouds (and the
temperature \tbgswarm\ of this dust).  These parameters also have
distinct effects on the mid-infrared colours $\ltwelve/\ltwofive$, 
$\lseven/\lfifteen$ and $\lfifteen/ \lsixty$, indicating that they can be 
constrained independently from fits of extended infrared spectral 
energy distributions. The mid-infrared colours are primarily controlled
by the different components of hot dust: the relative contributions 
\xipahs\ and \ximir\ of PAHs and the hot mid-infrared continuum to the
the infrared luminosity of stellar birth clouds, and the contribution
by PAHs to the infrared luminosity of the ambient ISM, which is controlled
indirectly by \xicold. In Section~\ref{extraction} below, we show how well
these various model parameters can be constrained in galaxies with observed
infrared spectral energy distributions.

\begin{table}
\centering
\caption{Parameters of the `standard', `cold' and `hot' models plotted in 
Figs.~\ref{varypar1} and \ref{varypar2}.}
\begin{tabular}{lccc}
\hline
Model & `cold' & `standard' & `hot' \\
\hline
\fmu        & 0.75 & 0.60 & 0.20 \\
\\[-5pt]
\xicold    & 0.75 & 0.80 & 0.90 \\
\\[-5pt]
\xipahs     & 0.45 & 0.05 & 0.01 \\
\\[-5pt]
\ximir       & 0.15 & 0.15 & 0.09 \\
\\[-5pt]
\xiwarm   & 0.40 & 0.80 & 0.95  \\
\\[-5pt]
\tbgswarm \  $\,\,$(K) &  40   & 48   & 55   \\
\\[-5pt]
\tbgscold \ (K) &  18   & 22   &  25   \\
\hline
\end{tabular}
\label{table_par}
\end{table}

\subsubsection{Constraints on dust mass}
\label{dust_mass}

It is also of interest to derive constraints on the dust mass in galaxies.
The mass $M_{\mathrm d}(T_{\mathrm d})$ in dust grains in thermal equilibrium
at the temperature $T_{\mathrm{d}}$ can be estimated from the far-infrared
radiation $L_\lambda^{T_{\mathrm{d}}}$  of these grains using the formula
\citep{HIL83}
\begin{equation}
L_\lambda^{T_{\mathrm{d}}}=
4\pi\,M_{\mathrm d}(T_{\mathrm d})\,\kappa_\lambda\,B_{\lambda}
(T_\mathrm{d})\,,
\label{mdust}
\end{equation}
where $\kappa_\lambda$ and $B_{\lambda}$ have been defined before 
(equation~\ref{mbb}). We adopt this formula to estimate the mass 
contributed by dust in thermal equilibrium in stellar birth clouds
(with temperature $T_{\mathrm{d}}= \tbgswarm$) and in the ambient 
ISM (with temperatures $T_{\mathrm{d}}= \tbgswarmism$ and 
$T_{\mathrm{d}}=\tbgscold$ ) in our model. We adopt as before
a dust emissivity index $\beta=1.5$ for warm dust and $\beta=2$
for cold dust and normalise $\kappa_\lambda$ at 850\,\mic\ assuming 
$\kappa_{850 \mu m}= 0.77\,$g$^{-1}$cm$^{2}$ \citep{DUNNE00}. Using 
equations~(\ref{mbb}), (\ref{ldust_bc2}), (\ref{ldust_ism2}) and
(\ref{mdust}), we compute the mass contributed by warm dust in stellar
birth clouds and in the ambient ISM as
\begin{equation}
\mwarmbc=\xiwarm\,(1-\fmu)\,\ldust\,
\left[
4\pi\,\int_0^\infty d\lambda\,\kappa_\lambda\,B_{\lambda}(\tbgswarm)
\right] ^{-1} 
\label{md_warmbc}
\end{equation}
and
\begin{equation}
\mwarmism=\xiwarmism\,\fmu\,\ldust\,
\left[
4\pi\,\int_0^\infty d\lambda\,\kappa_\lambda\,B_{\lambda}(\tbgswarmism)
\right] ^{-1} \,,
\label{md_warmism}
\end{equation}
and that contributed by cold dust in the ambient ISM as
\begin{equation}
\mcoldism=\xicold\,\fmu\,\ldust\,
\left[
4\pi\,\int_0^\infty d\lambda\,\kappa_\lambda\,B_{\lambda}(\tbgscold)
\right] ^{-1} \,.
\label{md_ism}
\end{equation}

To include the contribution by stochastically heated dust grains
(not in thermal equilibrium; see Section~\ref{components}), we adopt
a standard \cite{MRN77} distribution of grain sizes $N(a)\propto 
a^{-3.5}$ over the range $0.005\,\mic\leq a\leq0.25\,\mic$.
We assume that the stochastically heated dust grains are very small
($a\leq0.01\,\mic$) and have mass densities typical of graphite, 
$\rm \rho\approx 2.26\,g\,cm^{-3}$, while bigger grains in thermal
equilibrium have mass densities typical of silicates, $\rm \rho\approx
3.30\,g\,cm^{-3}$ \citep{DL84}. For these assumptions, the mass 
contributed by grains of all sizes is about 5 per cent larger
than that contributed by big grains alone ($a>0.01\,\mic$). The 
contribution by PAHs to the overall dust mass is also very small, of
the order of a few per cent \citep{DRAINE07}. We therefore estimate 
the total dust mass of a galaxy as
\begin{equation}
M_\mathrm{d}\approx1.1\,(\mwarmbc+\mwarmism+\mcoldism)\,,
\label{mdust2}
\end{equation}
where \mwarmbc, \mwarmism\ and \mcoldism\ are given by
equations~(\ref{md_warmbc}), (\ref{md_warmism}) and (\ref{md_ism}). 


	\subsection{Combined ultraviolet, optical and infrared spectral
        energy distributions}
	\label{combination}

A main feature of our model is the consistent modelling of ultraviolet,
optical and infrared spectral energy distributions of galaxies. This is
achieved by first computing the total energy absorbed by dust in stellar
birth clouds and in the ambient ISM, $L_{\mathrm{d}}^{\,\mathrm{BC}}$
and $L_{\mathrm{d}}^{\, \mathrm{ISM}}$ (Section~\ref{stellar}), and then
re-distributing it at infrared wavelengths (Section~\ref{irmodel}). The
main assumptions are the conservation of the energy absorbed and 
reradiated by dust, and that the dust in the ISM of galaxies is heated
only by starlight (in particular, we ignore the possible influence of
an active galactic nucleus). Different combinations of star formation 
histories, metallicities and dust contents can lead to the same absorbed
energies $L_{\mathrm{d}}^{\, \mathrm{BC}}$ and $L_{\mathrm{d}}^{\, 
\mathrm{ISM}}$ in a model galaxy. Furthermore, these energies can be 
distributed in wavelength using different combinations of dust parameters 
in the stellar birth clouds (\xipahs, \ximir, \xiwarm\ and \tbgswarm) and
the ambient ISM (\xicold\ and \tbgscold). In our model, therefore,
a wide range of ultraviolet and optical spectral energy distributions
can be associated to a wide range of infrared spectral energy 
distributions, at fixed $L_{\mathrm{d}}^{\, \mathrm{BC}}$ and
$L_{\mathrm{d}}^{\,\mathrm{ISM}}$ (or equivalently, at fixed \fmu\ 
and \ldust; see equations \ref{ldust} and \ref{fmu1}).  In 
Section~\ref{extraction} below, we show how combined observations at
ultraviolet, optical and infrared wavelengths can be used to uniquely 
constrain the star formation histories and dust properties of galaxies
using this model.

\begin{figure*}
\centering
\includegraphics[width=0.8\textwidth]{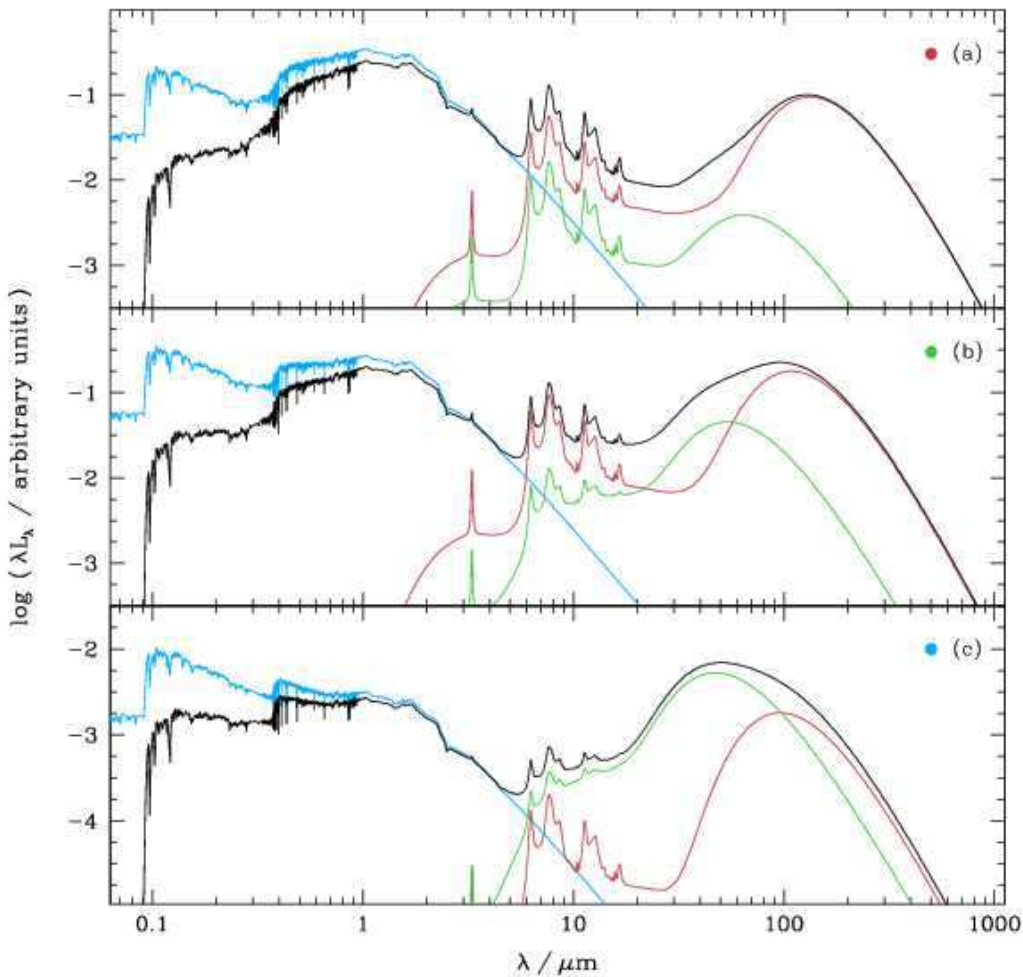}
\caption{Examples of spectral energy distributions obtained by combining
the infrared models of Table~\ref{table_par} with attenuated stellar 
population spectra corresponding to the same contributions by dust in
stellar birth clouds ($1-\fmu$) and in the ambient ISM (\fmu) to the total 
energy \ldust\ absorbed and reradiated by dust (Section~\ref{combination}).
(a) Quiescent star-forming galaxy spectrum combined with the `cold' 
infrared model of Table~\ref{table_par}; (b) normal star-forming galaxy
spectrum combined with the `standard' infrared model of
Table~\ref{table_par}; (c) starburst galaxy spectrum combined with the
`hot' infrared model of Table~\ref{table_par} (see text for details 
about the parameters of the stellar population models). Each panel 
shows the unattenuated stellar spectrum (blue line), the emission by dust
in stellar birth clouds (green line), the emission by dust in the ambient
ISM (red line) and the total emission from the galaxy, corresponding to 
the sum of the attenuated stellar spectrum and the total infrared emission
(black line).}
\label{example_seds2}
\end{figure*}

To illustrate the combination of ultraviolet, optical and infrared
spectral energy distributions with our model, we compute examples of
attenuated stellar population spectra consistent with the \fmu\ 
parameters of the cold, standard and hot infrared models of
Table~\ref{table_par} (Section \ref{colours}). For simplicity, we 
assume solar metallicity and select models with exponentially declining
star formation rates
\begin{equation}
\psi(t) \propto \exp({-\gamma\,t})\,,
\label{sfr}
\end{equation}
where $\gamma$ is the star formation timescale parameter. We choose
models with $\gamma=0$, 0.07 and 0.25~Gyr$^{-1}$ at the ages $t=1.4$,
10 and 10~Gyr, respectively, to represent a starburst, a normal 
star-forming and a quiescent star-forming galaxy (e.g. \citealt{KC98}).
For the attenuation of starlight by dust (equations \ref{tau_bc} and 
\ref{tau_ism}), we adopt effective dust absorption optical depths
$\hat \tau_{V}=2.0$, 1.5 and 1.0, for the starburst, normal 
star-forming and quiescent star-forming models, respectively. These
values are consistent with the expectation that more actively 
star-forming galaxies are more obscured (e.g. \citealt{WH96,HOPK01,
SULL01}). We adopt fractions $\mu=0.1$, 0.3 and 0.5 of $\hat \tau_{V}$
arising from dust in the ambient ISM. The resulting model galaxies
have $\fmu\approx0.2$, 0.6 and 0.75, respectively, consistent with
the parameters of the hot, standard and cold infrared models of 
Table~\ref{table_par}.

In Fig.~\ref{example_seds2}, we show the combined spectral energy 
distributions of the starburst + hot infrared, normal star-forming
+ standard infrared, and quiescent star-forming + cold infrared models,
after scaling in each case the infrared luminosity to the total 
luminosity \ldust\ absorbed by dust. As we shall see in the next section,
the ability to compute such combined ultraviolet, optical and infrared 
spectral energy distributions for wide ranges of physical parameters
of galaxies has important implications for statistical estimates of 
star formation histories and dust properties.


\section{Constraints on physical parameters from multi-wavelength galaxy 
observations}
\label{extraction}

In this section, we show how the simple model described in 
Section~\ref{the_model} can be used to extract star formation 
histories and dust properties from ultraviolet, optical and infrared
observations of galaxies. In Section~\ref{approach}, we first describe
our methodology to derive statistical constraints on galaxy physical
parameters from multi-wavelength observations. Then, in 
Section~\ref{application}, we use this methodology to constrain the
physical parameters of star-forming galaxies in the SINGS sample. We
compare our results to those that would be obtained using previous 
models in Section~\ref{comparison_models}. Finally, we discuss
possible sources of systematic errors associated with our approach
in Section~\ref{systematic_errors} and summarize the applicability 
of this model in Section~\ref{applicability}.


	\subsection{Methodology}
	\label{approach}

The model of Section~\ref{the_model} allows one to compute the 
ultraviolet, optical and infrared emission from galaxies.
This model contains a minimum number of adjustable parameters 
required to account for the observed relations between various integrated 
spectral properties of galaxies: age, star formation history, stellar 
metallicity, two components for the attenuation by dust and four
contributors to the infrared emission (PAHs, hot mid-infrared continuum,
warm and cold dust in thermal equilibrium). The large number of observable
quantities to which the model can be compared insures that these
different adjustable parameters can be constrained in a meaningful 
way (see below). A usual limitation of this type of study is that
several different combinations of physical parameters can lead to 
similar spectral energy distributions of galaxies. For example, age,
metallicity and dust attenuation have similar effects on the 
ultraviolet and optical spectra of galaxies. An efficient way to
derive statistical constraints on the various parameters in these
conditions is to consider a wide library of models encompassing all
plausible parameter combinations. Given an observed galaxy, we can
build the likelihood distribution of any physical parameter by 
evaluating how well each model in the library can account for the
observed properties of the galaxy. This Bayesian approach is similar
to that used, for example, by \cite{KAUF03} to interpret the optical
spectra of SDSS galaxies. The underlying assumption is that the 
library of models is the distribution from which the data were 
randomly drawn. Thus, the prior distribution of models must be such
that the entire observational space is reasonably well sampled, and
that no a priori implausible corner of parameter space accounts for
a large fraction of the models.

	\subsubsection{Model library}
	\label{library}

We build a comprehensive library of models by generating separately
a random library of stellar population models, for wide ranges of
star formation histories, metallicities and dust contents, and a
random library of infrared spectra, for wide ranges of dust temperatures
and fractional contributions by the different dust components to the
total infrared luminosity. We then combine these libraries following
the procedure outline in Section~\ref{combination}.

For simplicity, we follow \cite{KAUF03} and parametrise each star 
formation history in the stellar population library in terms 
of two components: an underlying continuous model, characterised
by an age \tg\ and a star formation time-scale parameter 
$\gamma$ (equation~\ref{sfr}), and random bursts superimposed on 
this continuous model. We take \tg\ to be uniformly distributed
over the interval from 0.1 and 13.5~Gyr. To avoid oversampling
galaxies with negligible current star formation, we distribute
$\gamma$ using the probability density function $p(\gamma)=1-
\tanh(8\,\gamma-6)$, which is approximately uniform over the interval
from 0 to 0.6~Gyr$^{-1}$ and drops exponentially to zero around
$\gamma=1$~Gyr$^{-1}$. Random bursts occur with equal probability at
all times until \tg. We set the probability so that 50 per cent of
the galaxies in the library have experienced a burst in the past 2~Gyr.
We parametrise the amplitude of each burst as 
$A=M_{\mathrm{burst}}/M_{\mathrm{cont}}$, where $M_{\mathrm{burst}}$
is the mass of stars formed in the burst and $M_{\mathrm{cont}}$ is
the total mass of stars formed by the continuous model over the time
\tg. This ratio is distributed logarithmically between 0.03 and 4.0.
During a burst, stars form at a constant rate over the time 
$t_{\mathrm{burst}}$, which we distribute uniformly between $3 
\times 10^7$ and $3 \times 10^8$~yr. We distribute the models 
uniformly in metallicity between 0.02 and 2 times solar.

We sample attenuation by dust in the library by randomly drawing 
the total effective {\it V}-band absorption optical depth, \tauv,
and the fraction of this contributed by dust in the ambient ISM,
$\mu$ (equations \ref{tau_bc} and \ref{tau_ism}). We distribute \tauv\
according to the probability density function $p(\tauv)=1-\tanh(1.5\,
\tauv-6.7)$, which is approximately uniform over the interval from 0
to 4 and drops exponentially to zero around $\tauv=6$. For $\mu$, we
adopt the same probability density function as for $\gamma$ above, 
i.e. $p(\mu)=1-\tanh(8\,\mu-6)$. We note that these priors for
attenuation encompass the dust properties of SDSS galaxies, for which
\tauv\ and $\mu$ peak around 1.0 and 0.3, respectively, with broad
scatter \citep{JB04,KONG04}. Our final stellar population
library consists of 50,000 different models.

In parallel, we generate a random library of infrared spectra as
follows. We take the fraction \fmu\ of the total infrared luminosity 
contributed by dust in the ambient ISM to be uniformly distributed
over the interval from 0 to 1. We adopt a similar distribution for 
the fractional contribution by warm dust in thermal equilibrium
to the infrared luminosity of stellar birth clouds, \xiwarm.
For each random drawing of \xiwarm, we successively draw the 
contributions by the other dust components to the infrared luminosity
of stellar birth clouds (i.e., hot mid-infrared continuum and 
PAHs) to satisfy the condition in equation~(\ref{xisumbc}): we draw
\ximir\ from a uniform distribution between 0 and $1-\xi_w^\mathrm{\,BC}$,
and we set $\xi_\mathrm{PAH}^\mathrm{\,BC}=1-\xi_w^\mathrm{\,BC}-\xi_
\mathrm{MIR}^{\,BC}$. While this procedure does not exclude values of
\ximir\ and \xipahs\ close to unity, it does favour small values of 
these parameters, and hence, it avoids oversampling physically 
implausible models. We take the equilibrium temperature \tbgswarm\ of
warm dust in the stellar birth clouds to be uniformly distributed 
between 30 and 60~K, and that \tbgscold\ of cold dust in the ambient
ISM to be unifomly distributed between 15 and 25~K. We draw the 
fractional contribution \xicold\ by cold dust in thermal equilibrium
to the infrared luminosity of the ambient ISM from a uniform distribution
between 0.5 and 1 (this also defines the contributions \xipahsism, 
\ximirism\ and \xiwarmism\ by PAHs, the hot mid-infrared continuum and warm
dust to the infrared luminosity of the ambient ISM, as described in 
Section~\ref{components}). Our final library of infrared spectra consists
of 50,000 different models.

We combine the library of stellar population models and that of 
infrared spectra by associating together models with similar \fmu,
which we scale according to the total infrared luminosity \ldust,
as outlined in Section~\ref{combination}. In practice, we associate
each model in the stellar population library to all the models in 
the infrared spectral library that have similar \fmu\ to within some
error interval $\delta f_{\mu}$.\footnote{This allows us to include
the uncertainties that could arise, for example, from orientation 
effects, in the connection between the stellar and dust emission.}
That is, we associate each stellar population spectrum, characterised
by $\fmu^{\mathrm{SFH}}$, to all the infrared spectra characterised
by  $\fmu^{\mathrm{IR}}$, such that \fmu $^{\mathrm{IR}}= 
f_{\mu}^{\mathrm{SFH}} \pm \delta f_{\mu}$. Each spectral 
combination satisfying this condition is included in the final 
model library and is assigned a value $\fmu = (\fmu^{\mathrm{SFH}}+
\fmu^{\mathrm{IR}})/2$. We adopt $\delta f_{\mu}=0.15$, which allows
good reproductions of combined ultraviolet, optical and infrared 
observations of galaxies (Section \ref{application}).

Our final library of combined ultraviolet, optical and infrared
spectral energy distributions consists of about 661 million models.
It is important to note that such a large number of models is required
to properly sample the multi-dimensional observational space. 

	\subsubsection{Statistical constraints on physical parameters}
	\label{estimation}

We now investigate the accuracy to which the model library described in
the previous section can help us constrain the star formation histories
and dust properties of galaxies for which multi-wavelength observations
are available. To assess this, we evaluate how well we can recover the
parameters of a random set of models with known properties, based on 
spectral fits. We consider the following observable quantities: the {\it 
GALEX} far-ultraviolet ({\it FUV}, 1520 \AA) and near-ultraviolet ({\it
NUV}, 2310 \AA) luminosities; the optical {\it UBV} luminosities; the
2MASS near-infrared {\it JHKs} (1.25, 1.65 and 2.17 $\mu$m) 
luminosities; the {\it Spitzer}/IRAC 3.6, 4.5, 5.8 and 8.0~\mic\ 
luminosities; the {\it ISO}/ISOCAM 6.75 and 15~\mic\ luminosities; 
the {\it IRAS} 12, 25, 60 and 100~\mic\ luminosities; the {\it 
Spitzer}/MIPS \citep{MIPS} 24, 70 and 160~\mic\ luminosities;
and the SCUBA \citep{SCUBA} 850~\mic\ luminosity. We also
include the hydrogen H$\alpha$ and H$\beta$ recombination-line luminosities.

\begin{figure*}
\centering
\includegraphics[width=0.8\textwidth]{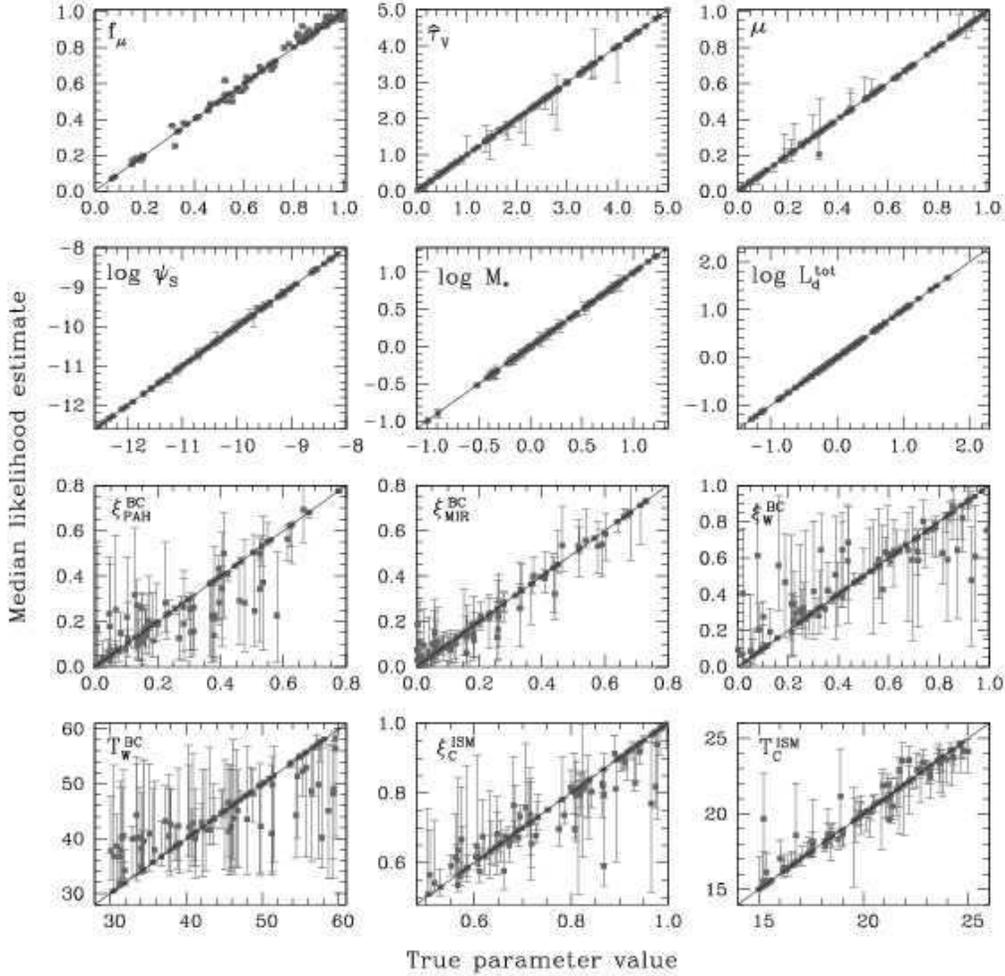}
\caption{Median-likelihood estimates of 12 parameters (indicated in 
the upper left corner of each panel) recovered from the spectral fits
of 100 mock galaxies, compared to the true values of these parameters
(see Section~\ref{estimation}). The error bars indicate the 16--84 
percentile ranges in the recovered probability distributions of the
parameters. These results were obtained by fitting simulated {\it GALEX}
({\it FUV} and {\it NUV}), optical ({\it UBV}), 2MASS ({\it JHKs}),
{\it Spitzer}/IRAC (3.6, 4.5, 5.8 and 8.0~\mic), {\it ISO}/ISOCAM 
(6.75 and 15\mic), {\it IRAS} (12, 25, 60, 100~\mic), {\it 
Spitzer}/MIPS (24, 70, and 160~\mic), SCUBA (850~\mic),
and H$\alpha$ and H$\beta$
luminosities.}
\label{estimate}
\end{figure*}

We compute these observable quantities from the spectra of all 661
million models in the library and randomly select a subset of 100 
models. To mimic observational conditions, we perturb the luminosities
of this subset of models assuming a fixed uncertainty of 10 per cent.
We then perform spectral fits to recover the likelihood distributions
of the physical parameters of these `mock galaxies' as follows. We 
first compare the luminosities of each mock galaxy in the sample to 
the luminosities of every model $j$ in the library to measure the 
$\chi^2$ goodness-of-fit of that model (e.g., \citealt{BEV03})
\begin{equation}
\chi_{j}^{2}=\sum_{i}^{} \left(\frac{L_{\nu}^{i}- w_j 
\times L_{\nu,j}^{i}}{\sigma_{i}}\right)^{2}\,,
\label{chi2}
\end{equation}
where $L_{\nu}^{i}$ and $L_{\nu,j}^{i}$ are the luminosities in the 
$i^{\rm th}$ band of the mock galaxy and the $j^{\rm th}$ model, 
respectively, $\sigma_{i}$ is the (10 percent) uncertainty in 
$L_{\nu}^{i}$, and
\begin{equation}
\displaystyle
w_j=\left(\sum_{i}^{} \frac{L_\nu^{i}L_{\nu,j}^{i}}
{\sigma_i^2}\right)
\left[\sum_{i}^{}
\left(\frac{L_{\nu,j}^{i}}{\sigma_i}\right)^2\right]^{-1}
\end{equation}
is the model scaling factor that minimizes $\chi_{j}^2$. Then, we build
the probability density function of any physical parameter of the mock
galaxy by weighting the value of that parameter in the $j^{\rm th}$ model
by the probability $\exp({-\chi_{j}^{2}/2})$. We take our `best estimate'
of the parameter to be the median of the resulting probability density 
function and the associated confidence interval to be the 16--84 
percentile range (this would be equivalent to the $\pm\,1 \sigma$ range
in the case of a Gaussian distribution).

In Fig.~\ref{estimate}, we compare the median-likelihood estimates to
the true values of 12 parameters recovered in this way for the 100
mock galaxies in our sample. These parameters are: the fraction of
total infrared luminosity contributed by dust in the ambient ISM, \fmu;
the total effective $V$-band absorption optical depth of the dust, $\hat
\tau{_V}$; the fraction of this contributed by dust in the ambient ISM,
$\mu$; the specific star formation rate, \ssfr, defined as the ratio
\begin{equation}
\ssfr(t)=\frac{\int_{t-t_8}^t dt'\,\psi(t')}{t_8\,M_\ast}
\label{ssfr_def}
\end{equation}
of the star formation rate averaged over the past $t_8=10^8$yr to the
current stellar mass $M_\ast$ of the galaxy; the stellar mass, $M_\ast$
(this accounts for the mass returned to the ISM by evolved stars); the 
total infrared luminosity, \ldust; the fractional contributions
by PAHs, the hot mid-infrared continuum and warm dust in thermal
equilibrium to the infrared luminosity of stellar birth clouds,
\xipahs, \ximir\ and \xiwarm; the equilibrium temperature of warm
grains in stellar birth clouds, \tbgswarm; the contribution by cold
dust to the total infrared luminosity of the ambient ISM, \xicold, 
and the equilibrium temperature of this dust, \tbgscold. Most of these 
parameters are recovered remarkably well by our model. In particular,
\fmu, $\hat\tau_V$, $\mu$, \ssfr, $M_\ast$, \ldust, \ximir, \xicold\
and \tbgscold\ are recovered to very high accuracy. The most uncertain
parameters are \xipahs, \xiwarm\ and \tbgswarm, for which the typical
confidence intervals reach almost 0.14, 0.18 and 11~K, respectively.
We conclude that our model is a valuable tool for deriving statistical
constraints on the star formation histories and dust properties of 
galaxies for which multi-wavelength (ultraviolet, optical and infrared)
observations are available.


	\subsection{Application to an observational sample}
	\label{application}

	\subsubsection{The sample}
	\label{sample}

Here, we exploit our model to interpret a wide range of ultraviolet,
optical and infrared observations of a sample of well-studied
nearby galaxies: the Spitzer Infrared Nearby Galaxy Survey (SINGS;
\citealt{K03}).  This sample contains 75 galaxies at a median
distance of 9.5 Mpc (we adopt a Hubble constant $\rm H_0=70\,km\,
s^{-1}Mpc^{-1}$). The galaxies span wide ranges in morphology (from
E-S0 to Im-I0) and star formation activity (from quiescent to starburst).
Some galaxies include low-luminosity active galactic nuclei 
(AGNs).\footnote{The low-luminosity AGNs should have a negligible
impact on the integrated broad-band fluxes of the galaxies. We 
distinguish galaxies hosting low-luminosity AGNs from normal 
star-forming galaxies in the spectral analyses presented later in
this paper.} We note that this sample extends out to lower total 
infrared luminosities than the \cite{DE02} sample used to
calibrate the infrared properties of our model in Section~\ref{colours}
($\ldust\la 10^{11} L_\odot$ instead of $\ldust\la 10^{12}L_\odot$), 
but it includes observations across a much wider range of wavelengths.

Observations at ultraviolet, optical and infrared wavelengths are
available for most galaxies in this sample. {\it GALEX} ultraviolet
({\it NUV} and {\it FUV}) observations are available for 70 galaxies
\citep{DALE07}. In the optical, we adopt {\it UBV} fluxes from the
RC3 for 65 galaxies \citep{VAUC91}.\footnote{We choose not
to use the optical {\it BVRI} fluxes tabulated by \cite{DALE07},
because of calibration problems in the optical photometry of the
SINGS Fifth Enhanced Data Release (D. Dale \& A. Gil de Paz, priv.
comm.)} Near-infrared {\it JHKs} fluxes are available for all galaxies
from the 2MASS Large Galaxy Atlas \citep{JARRETT03}. In the mid- and
far-infrared, we use the {\it Spitzer} observations published by
\cite{DALE07}. This includes IRAC photometry at 3.6, 4.5, 5.8 and
8.0~\mic\ and MIPS photometry at 24, 70 and 160~\mic. {\it ISO},
{\it IRAS} and SCUBA 850~\mic\ observations are also available for
13, 65 and 22 SINGS galaxies, respectively. These observations are
typically of lower quality than {\it Spitzer} data, with photometric
uncertainties around 20 per cent for {\it ISO} and {\it IRAS}
and 30 per cent for SCUBA. More detail about the photometry of
SINGS galaxies can be found in \cite{DALE05, DALE07}. 

\begin{figure}
\centering
\includegraphics[width=0.5\textwidth]{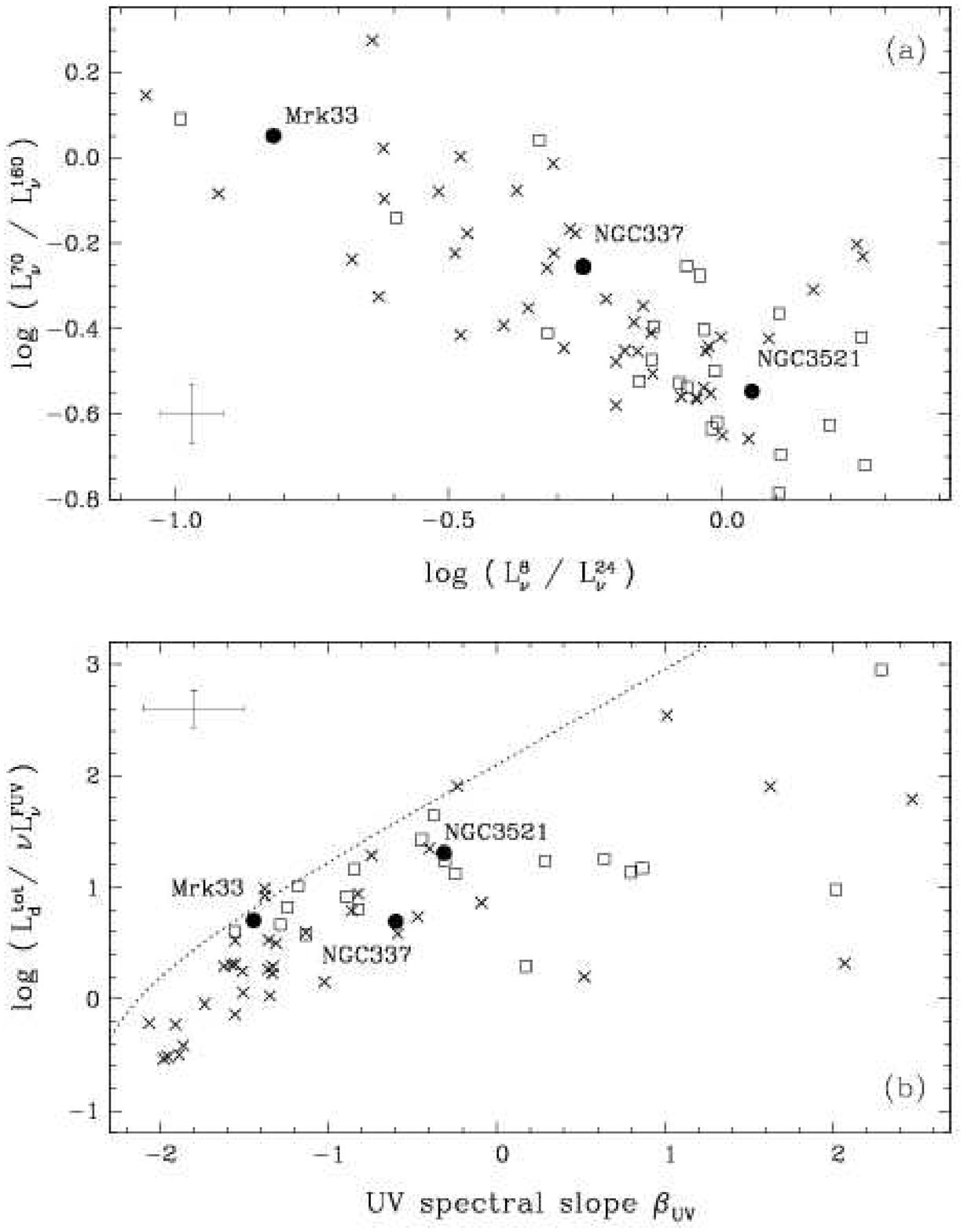}
\caption{Selected properties of the SINGS galaxy sample discussed in
Section~\ref{application}. (a) Ratio of 70-\mic\ to 160-\mic\ luminosity
plotted against ratio of 8-\mic\ to 24-\mic\ luminosity. (b) Ratio 
of total-infrared to ultraviolet luminosity as a function of 
ultraviolet spectral slope (the IRX--UV diagram). The dotted line 
in this diagram shows a fit by \protect\cite{KONG04} to the relation
followed by starburst galaxies \protect\citep{MEURER99}. In both
panels, crosses refer to galaxies with no AGN activity according to
\protect\cite{K03}, while open squares refer to galaxies hosting
low-luminosity AGNs (LINER and Seyfert). Filled circled indicate three
galaxies selected along the $\lseventy/\lonesixty$ versus $\leight/
\ltwofour$ relation, for which a wide set of observations are available 
(Section~\ref{results}). Typical measurement errors are indicated 
in each diagram.}
\label{irx_beta}
\end{figure}

It is important to note that these multi-wavelength observations
can be combined in a meaningful way, since the fluxes all pertain to
the total emission from a galaxy. For example, the {\it GALEX} 
({\it FUV} and {\it NUV}), RC3 ({\it UBV}) and 2MASS ({\it JHKs}) 
fluxes were obtained by integrating extrapolated surface brightness
profiles to include the entire emission from a galaxy (see
\citealt{VAUC91,JARRETT03,GdP06}). Also, the infrared {\it
Spitzer} IRAC and MIPS fluxes include extended-source aperture 
corrections (amounting typically to 10 per cent; see 
\citealt{DALE05,DALE07}). Following \cite{DRAINE07}, we exclude
9 galaxies with bad IRAC and MIPS detections (because of 
contamination by external sources; very low surface brightness 
compared to foreground cirrus emission; and saturation issues): 
NGC~584, NGC~3034, NGC~1404, NGC~4552, M81~DwarfA, M81~DwarfB, 
DDO~154, DDO~165 and Holmberg~IX. For 19 galaxies, we include
\ha\ and \hb\ emission-line fluxes (corrected for contamination by
stellar absorption) from the integrated spectroscopy of \cite{MK06}.
Our final sample includes 66 galaxies, of which 61 have {\it GALEX}
measurements. As before, for the purpose of comparisons with models,
we neglect possible anisotropies and equate flux ratios at all 
wavelengths to the corresponding luminosity ratios.

We illustrate the ultraviolet and infrared properties of the SINGS 
galaxies in Fig.~\ref{irx_beta}. Fig.~\ref{irx_beta}a shows the MIPS
$\lseventy/\lonesixty$ luminosity ratio plotted against the IRAC+MIPS
$\leight/\ltwofour$ luminosity ratio. SINGS galaxies in this diagram 
follow a sequence very similar to that followed by the galaxies of the 
\cite{DE02} sample of Section~\ref{irmodel} in the analogous {\it IRAS}
$\lsixty/ \lhundred$ versus $\ltwelve/\ltwofive$ diagram (Fig.~\ref{varypar1}).
In Fig.~\ref{irx_beta}b, we plot ratio of total-infrared to ultraviolet
luminosity $\ldust/L_{\mathrm FUV}$ as a function of ultraviolet spectral
slope $\beta_\mathrm{UV}$ (the `IRX-UV diagram'). We used equation (4)
of \cite{DH02} to estimate \ldust\ from the MIPS observations at 24, 70 
and 160~\mic, and equation (1) of \cite{KONG04} to compute 
$\beta_\mathrm{UV}$ from the {\it GALEX} {\it FUV} and {\it NUV} 
luminosities. Previous studies have shown that starburst galaxies 
follow a tight sequence in this diagram (indicated by the dotted line
in Fig.~\ref{irx_beta}b), while more quiescent star-forming galaxies 
tend to exhibit redder ultraviolet spectra (i.e. larger $\beta_\mathrm{
UV}$) than starburst galaxies at fixed $\ldust/L_{\mathrm FUV}$ 
\citep{MEURER99, BELL02,KONG04}. Fig.~\ref{irx_beta} confirms that the
SINGS galaxies are representative of the ultraviolet and infrared 
properties of nearby galaxies.

	\subsubsection{Example of constraints on physical parameters}
	\label{results}

\begin{figure*}
\centering
\includegraphics[width=11cm,height=11cm]{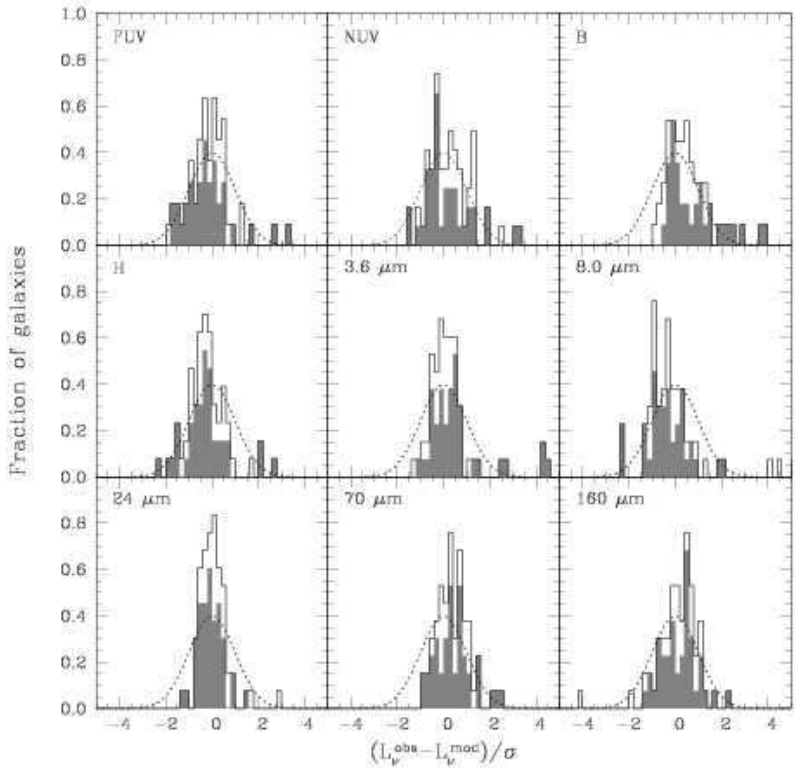}
\caption{Distribution of the difference between observed luminosity
$L_\nu^{\mathrm{obs}}$ and best-fit model luminosity $L_\nu^{\mathrm{mod}}$, 
in units of the observational error $\sigma^{\mathrm{obs}}$, for the
66 galaxies of the SINGS sample studied in Section~\ref{application}
Each panel refers to a different photometric band (as indicated). 
The best-fit model for each galaxy was selected by fitting as many
luminosities as available in the following bands: {\it GALEX} ({\it FUV} 
and {\it NUV}), optical ({\it UBV}), 2MASS ({\it JHKs}), {\it 
Spitzer}/IRAC (3.6, 4.5, 5.8 and 8.0~\mic), {\it ISO}/ISOCAM (6.75 and
15\mic), {\it IRAS} (12, 25, 60 and 100~\mic), {\it Spitzer}/MIPS (24, 70
and 160~\mic), SCUBA (850~\mic) and H$\alpha$ and
H$\beta$. In each panel, the plain
histogram shows the result for the full sample of 66 SINGS galaxies,
while the shaded histogram shows the result for the subsample of 44
galaxies with no AGN activity according to \protect\cite{K03}. The
dotted curve shows a Gaussian distribution with unit standard 
deviation, for reference.}
\label{residuals}
\end{figure*}

We now use the methodology outlined in Section~\ref{approach} to 
constrain the star formation histories and dust properties of the 
SINGS galaxies. We first assess how well our model can reproduce the
observed multi-wavelength observations of these galaxies. For each
observed galaxy, we select the model of the library presented 
in Section~\ref{approach} that minimizes $\chi^2_j$ (equation~\ref{chi2}),
as computed by including as many luminosities as available in the following
bands: {\it GALEX} ({\it FUV} and {\it NUV}), optical ({\it UBV}), 2MASS 
({\it JHKs}), {\it Spitzer}/IRAC (3.6, 4.5, 5.8 and 8.0~\mic), {\it 
ISO}/ISOCAM (6.75 and 15\mic), {\it IRAS} (12, 25, 60 and 100~\mic), 
{\it Spitzer}/MIPS (24, 70 and 160~\mic), SCUBA (850~\mic) and 
H$\alpha$ and H$\beta$. Fig.~\ref{residuals} shows the resulting 
distribution of the difference between the observed luminosity
$L_\nu^{\mathrm{obs}}$ and the best-fit model luminosity
$L_\nu^{\mathrm{mod}}$, in units of the observational error 
$\sigma^{\mathrm{obs}}$, for the 66 galaxies in the sample (plain
histograms). Each panel shows the result for a different photometric
band. For reference, the shaded histograms show the results obtained 
when excluding the 22 galaxies identified as hosts of low-luminosity 
AGNs by \citet{K03}. 

Fig.~\ref{residuals} shows that our model can reproduce simultaneously
the ultraviolet, optical and near-, mid- and far-infrared observations 
of all but a few SINGS galaxies to within the observational errors. 
This is remarkable, considering the relative simplicity of the model. We
find that the galaxies the least well fitted at ultraviolet and optical
wavelengths tend to be low-metallicity dwarf galaxies, such as the blue
compact dwarf galaxy NGC~2915 (e.g. \citealt{LEE03}) and the diffuse 
dwarf irregular galaxy Holmberg II (e.g. \citealt{HO98}). The difficulty
in reproducing these observations may arise either from a limitation of 
the spectral synthesis code, or of the simple dust prescription,
or from an underestimate of observational errors. The three galaxies for
which the models significantly underestimate the 3.6~\mic\ emission in
Fig.~\ref{residuals} are also low-metallicity dwarf galaxies (NGC4236,
IC4710 and NGC6822). For these galaxies, our simple scaling of the 
near-infrared continuum strength around 4~\mic\ with the flux density 
of the 7.7~\mic\ PAH feature may not be appropriate (see 
Section~\ref{components}). Fig.~\ref{residuals} further shows that
the galaxies with 8~\mic\ mid-infrared luminosities least well
reproduced by the model tend to be hosts of low-luminosity AGNs. In
these galaxies, the contribution to the infrared emission by dust heated
by the AGN may be significant. Despite these few outliers, we conclude
from Fig.~\ref{residuals} that our model is adequate to investigate the 
constraints set on the physical properties of SINGS galaxies by their
ultraviolet, optical and infrared spectral energy distributions.

To investigate the implications of these fits for the determination of
physical parameters, we focus on three galaxies spanning different dust
properties along the $\lseventy/\lonesixty$ versus $\leight/\ltwofour$
colour-colour relation in Fig.~\ref{irx_beta}a, for which a wide set of
observations are available: NGC~3521, a spiral (SABbc) galaxy at a distance
of 9 Mpc; NGC~337, a spiral (SBd) galaxy at a distance of 24.7 Mpc; and 
Mrk~33, a dwarf irregular (Im) starburst galaxy at a distance of 21.7 Mpc.
For all three galaxies, observations from {\it GALEX} ({\it FUV} and {\it
NUV}), 2MASS ({\it JHKs}), {\it Spitzer} (IRAC and MIPS), {\it IRAS}
and SCUBA
are available, along with integrated \ha\ and \hb\ spectroscopy from 
\cite{MK06}. For NGC~3521 and NGC~337, additional constraints are 
available from RC3 ({\it UBV}) and {\it ISO} (ISOCAM 6.75 and 15\mic)
photometry.

\begin{figure*}
\centering
\includegraphics[width=0.8\textwidth]{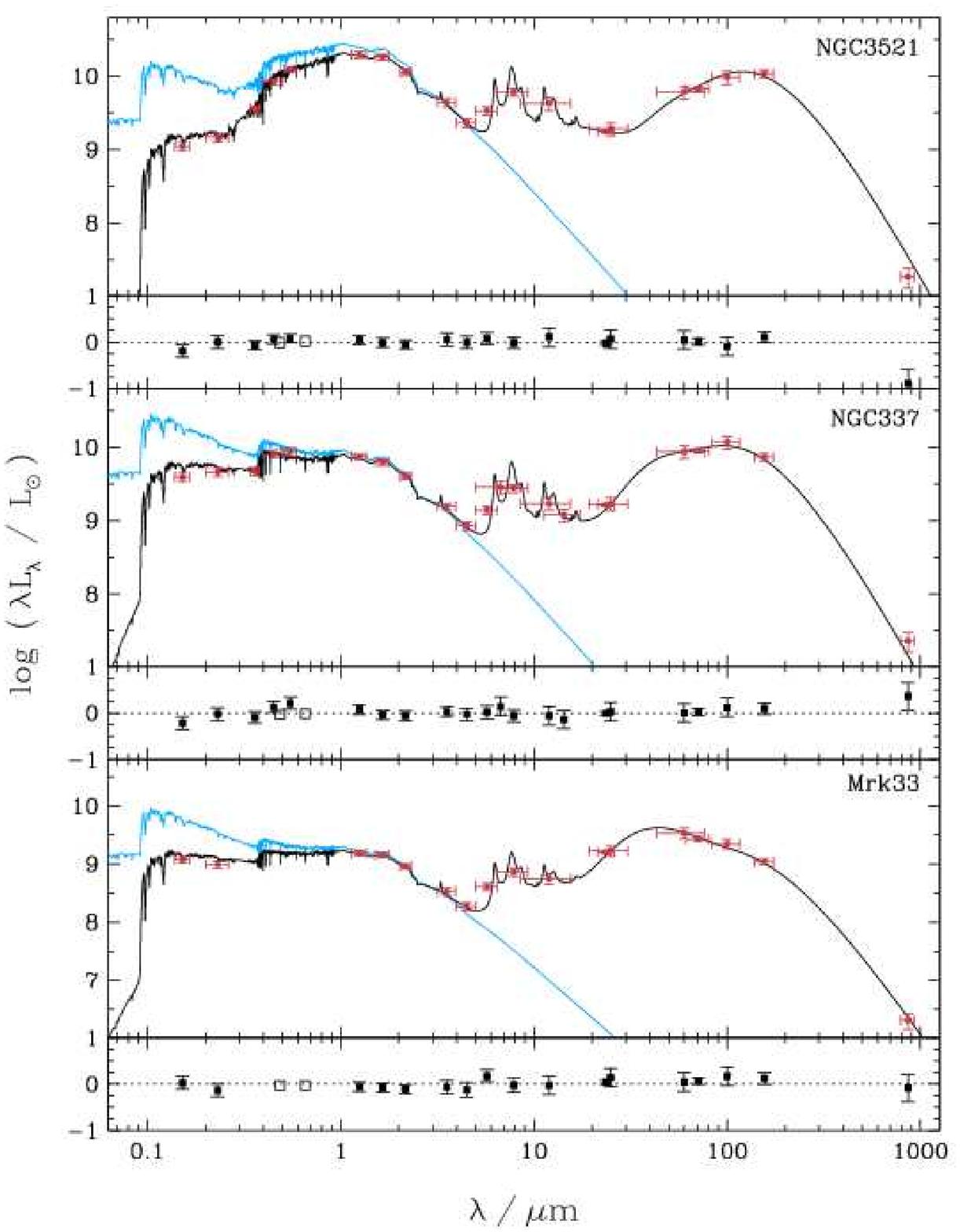}
\caption{Best model fits (in black) to the observed spectral energy 
distributions (in red) of the three galaxies NGC~3521 (top panel), 
NGC~3521 (middle panel) and Mrk~33 (bottom panel) spanning the infrared 
colour-colour relation of Fig.~\ref{irx_beta}a. In each panel, the 
blue line shows the unattenuated stellar population spectrum. For each
observational point, the vertical error bar indicates the measurement
error, while the horizontal error bar shows the effective width of the
corresponding photometric bands. The residuals 
$(L_\lambda^\mathrm{obs}-L_\lambda^\mathrm{mod})/
L_\lambda^\mathrm{obs}$ are shown at the 
bottom of each panel. Filled squares refer to broad-band luminosties
({\it GALEX} {\it FUV} and {\it NUV}; RC3 {\it UBV}; 2MASS {\it JHKs};
{\it Spitzer} IRAC and MIPS; {\it ISO}, {\it IRAS} and SCUBA); open squares
to the integrated \ha\ and \hb\ luminosities.}
\label{seds2}
\end{figure*}

Fig.~\ref{seds2} shows the models (in black) providing the best fits
to the observed ultraviolet, optical, \ha\ and \hb, and infrared 
spectral energy distributions of these three galaxies (in red).
The quality of the fits is remarkable, as quantified by the residuals
shown at the bottom of each spectrum. We note that, according to 
the best-fit models, the sequence of increasing $\fseventy/\fonesixty$
and decreasing $\feight/\ftwofour$ colours from NGC~3521 to NGC~3521 to
Mrk~33, which reflects a drop in the relative intensity of PAHs and
a blueshift of the peak infrared luminosity (i.e. a rise in the overall
dust temperature), is associated to an increase in star formation 
activity. This is apparent from the relative strengthening of the 
unattenuated ultraviolet and optical spectrum (in blue) from NGC~3521
to NGC~3521 to Mrk~33 in Fig.~\ref{seds2}. We investigate this trend
further in Section~\ref{statistics} below.

\begin{figure*}
\centering
\includegraphics[angle=270,width=\textwidth]{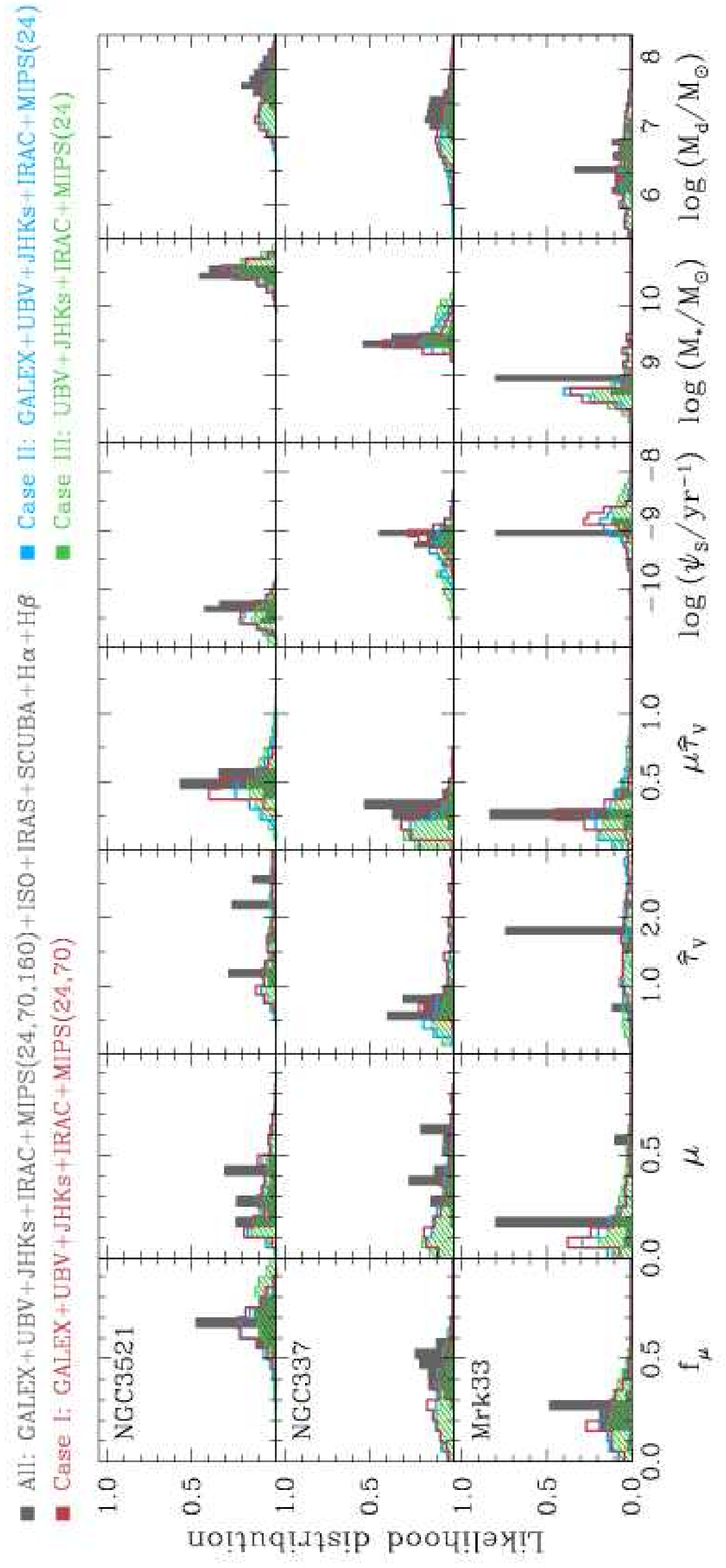}
\caption{Likelihood distributions of physical quantities derived from 
fits to the observed ultraviolet, optical and infrared spectral energy
distributions of NGC~3521 (top row), NGC~337 (middle row) and Mrk~33 
(bottom row): fraction of the total infrared luminosity contributed by 
dust in the ambient ISM ($f_\mu$); total effective $V$-band absorption
optical depth of the dust ($\hat\tau_V$); fraction of the total $V$-band
absorption optical depth of the dust contributed by the ambient ISM  ($\mu$); 
effective $V$-band absorption optical depth of the dust in the ambient
ISM ($\mu \hat\tau_V$); specific star formation rate ($\psi_S$); stellar
mass ($M_\ast$); and dust mass ($M_{\mathrm d}$). The different histograms
correspond to the different sets of observational constraints indicated at
the top of the figure (see text of Section~\ref{results} for more detail).}
\label{pdfs1}
\end{figure*}

\begin{figure*}
\centering
\includegraphics[angle=270,width=\textwidth]{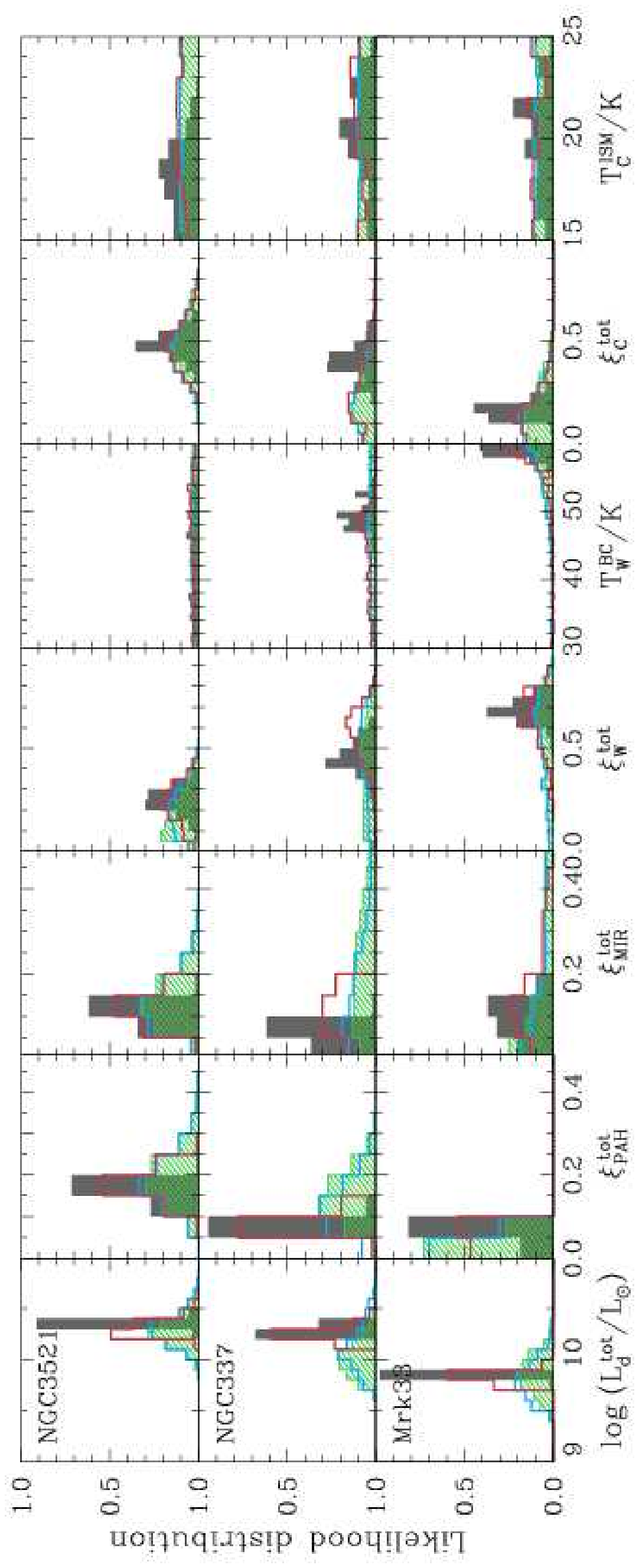}
\caption{Likelihood distributions of physical quantities derived from
fits to the observed ultraviolet, optical and infrared spectral energy
distributions of NGC~3521 (top row), NGC~337 (middle row) and Mrk~33
(bottom row): total infrared luminosity of the dust (\ldust); global
contributions (i.e. including stellar birth clouds and the ambient ISM)
by PAHs (\xipahstot), the hot mid-infrared continuum (\ximirtot) and warm 
dust in thermal equilibrium (\xiwarmtot) to the total infrared luminosity;
equilibrium temperature of warm dust in stellar birth clouds (\tbgswarm);
contribution by cold dust in thermal equilibrium to the total infrared
luminosity (\xicoldtot); and equilibrium temperature of cold dust in the
ambient ISM (\tbgscold). The different histograms correspond to the 
different sets of observational constraints indicated at the top of 
Fig.~\ref{pdfs1} (see text of Section~\ref{results} for more detail).}
\label{pdfs2}
\end{figure*}

We can study in more detail the constraints set on the star formation
histories and dust properties of the galaxies by looking at the 
probability distributions of the corresponding model parameters. 
In Figs.~\ref{pdfs1} and \ref{pdfs2}, we show the likelihood distributions 
of 14 quantities constrained by the observed spectral energy distributions 
of NGC~3521, NGC~337 and Mrk~33. These distributions were derived
using the model library of Section~\ref{library}, following the procedure
outlined in Section~\ref{estimation}. Fig.~\ref{pdfs1} shows the likelihood 
distributions of parameters pertaining to the star formation history and
dust content: fraction of the total infrared luminosity contributed by
dust in the ambient ISM ($f_\mu$); fraction of the total effective $V$-band
absorption optical depth of the dust contributed by the ambient ISM ($\mu$);
effective $V$-band absorption optical depth of the dust ($\hat\tau_V$); 
effective $V$-band absorption optical depth of the dust in the ambient 
ISM ($\mu \hat\tau_V$); specific star formation rate ($\psi_S$; 
equation~\ref{ssfr_def}); stellar mass ($M_\ast$); and dust mass 
($M_{\mathrm d}$; equation \ref{mdust2}). Fig.~\ref{pdfs2} shows the
likelihood distributions of parameters pertaining to the dust emission:
total infrared luminosity (\ldust); global contributions (i.e. including
stellar birth clouds and the ambient ISM) by PAHs (\xipahstot), the hot
mid-infrared continuum (\ximirtot) and warm dust in thermal equilibrium
(\xiwarmtot) to the total infrared luminosity (equations~\ref{exipahstot},
\ref{eximirtot} and \ref{exiwarmtot}); equilibrium temperature of warm dust
in stellar birth clouds (\tbgswarm); contribution by cold dust in thermal
equilibrium to the total infrared luminosity (\xicoldtot, 
equation~\ref{exicoldtot}); and equilibrium temperature of cold dust in
the ambient ISM (\tbgscold).

The different histograms in Figs.~\ref{pdfs1} and \ref{pdfs2} refer to 
different assumptions about the available set of observations. The grey
shaded histograms show the results obtained when using all the 
observational measurements from Fig.~\ref{seds2}. In this case,
all the parameters are well constrained by the observed ultraviolet,
optical and infrared spectral energy distributions of the galaxies, 
except for the equilibrium temperature \tbgswarm\ of warm dust in 
stellar birth clouds in NGC~3521. This is not surprising,
since the total infrared emission from this galaxy is largely dominated
by dust in the ambient ISM (as can be appreciated, for example, from 
the analogous spectrum of the quiescent star-forming galaxy of 
Fig.~\ref{example_seds2}a). We find that, as anticipated from 
Fig.~\ref{seds2}, the transition from NGC~3521 to NGC~337 to Mrk~33 
corresponds to a combined rise in specific star formation rate 
(\ssfr) and dust temperature (\tbgswarm\ and \tbgscold), a larger
contribution to the total infrared luminosity by warm dust (\xiwarmtot)
relative to cold dust (\xicoldtot), and by stellar birth clouds relative
to the ambient ISM (\fmu), and a weakening of PAH features (\xipahstot).
These results are consistent with the finding by \cite{DRAINE07}, on the
basis of a more sophisticated physical dust model, that the intensity of
the dust-heating starlight increases, and the fraction of dust mass
contributed by PAHs decreases, from NGC~3521 to NGC~337 to Mrk~33. We
emphasise the importance of finding tight constraints on all the 
adjustable model parameters in Figs.~\ref{pdfs1} and \ref{pdfs2}. This
confirms that the problem is well defined and that our model can be used
to derive meaningful constraints on the star formation histories and dust
properties of galaxies from multi-wavelength observations. 

In many situations, observations may not be available across the full
range from ultraviolet to far-infrared wavelengths to constrain the
physical parameters of galaxies. It is useful, therefore, to explore 
the constraints that can be derived on the physical parameters of
Figs.~\ref{pdfs1} and \ref{pdfs2} when reducing the set of available
observations. We illustrate this by considering three potentially
common situations:

\begin{enumerate}

\item{\it Case I} (red histograms in Figs.~\ref{pdfs1} and \ref{pdfs2}):
we relax the constraints on the \ha\ and \hb, {\it ISO}/ISOCAM (6.75 and
15~\mic), {\it IRAS} (12, 25, 60 and 100~\mic), {\it Spitzer}/MIPS
160~\mic\ and SCUBA 850~\mic\ luminosities. This leaves constraints from
{\it GALEX} ({\it FUV} and {\it NUV}), RC3 ({\it UBV}), 2MASS ({\it JHKs}),
{\it Spitzer}/IRAC (3.6, 4.5, 5.8 and 8.0~\mic) and {\it Spitzer}/MIPS (24
and 70~\mic).

The most important effect of excluding the constraints on $L_\mathrm{H 
\alpha}$ and $L_\mathrm{H\beta}$ is to lose the determination of the
total effective $V$-band absorption optical depth \tauv\ in Fig.~\ref{pdfs1}.
This is because emission lines are the best tracers of dust in stellar birth
clouds. In fact, the product $\mu\tauv$, i.e. the part of \tauv\ arising from
dust in the ambient ISM, remains reasonably well constrained by ultraviolet,
optical and infrared photometry alone. In Fig.~\ref{pdfs2}, the loss of 
far-infrared information redward of 70~\mic\ ({\it IRAS} 100 \mic, MIPS 
160 \mic\ and SCUBA 850 \mic) has the most dramatic effect on determinations
of the contribution by cold dust to the total infrared luminosity 
(\xicoldtot) and on the equilibrium temperature of both cold (\tbgscold)
and warm (\tbgswarm) dust. The IRAC data at wavelengths 3.6, 4.5, 5.8 and
8.0~\mic\ still provide valuable constraints on \xipahstot, while the loss
of {\it ISO} 6.75 and 15~\mic\ and {\it IRAS} 12~\mic\ information makes
determinations of \ximirtot\ and \xiwarmtot\ more uncertain. The lack of 
far-infrared information also weakens the constraints on the dust mass 
$M_{\mathrm d}$ in Fig.~\ref{pdfs1}.

\item{\it Case II} (blue histograms in Figs.~\ref{pdfs1} and \ref{pdfs2}):
this is similar to Case~I, but we also relax the constraint on the {\it 
Spitzer}/MIPS 70-\mic\ luminosity. The main effect is to significantly 
worsen estimates of the contributions to the total infrared luminosity
by warm dust in thermal equilibrium (\xiwarmtot) and by the hot
mid-infrared continuum (\ximirtot). The constraints on the total infrared
luminosity \ldust\ itself are also slightly worse.

\item{\it Case III} (green histograms in Figs.~\ref{pdfs1} and \ref{pdfs2}):
this is similar to Case II, but we also relax the constraints on the {\it
GALEX} {\it FUV} and {\it NUV} luminosities. Not including information about
the ultraviolet radiation from young stars has only a weak influence on the
results of Figs.~\ref{pdfs1} and \ref{pdfs2}, when optical, near-infrared 
and mid-infrared data out to 24~\mic\ are already available. The constraints
on the specific star formation rate (\ssfr), the attenuation parameters
($\mu$, \tauv\ and the product $\mu\tauv$) and the relative contribution 
by dust in the ambient ISM to the total infrared luminosity (\fmu) become
marginally worse. The total infrared luminosity \ldust\ and the stellar 
mass $M_\ast$ remain reasonably well constrained.

\end{enumerate}  

We note that, if a galaxy has mid-infrared colours characteristic
of Galactic cirrus emission, and if no other spectral information is 
available at shorter and longer wavelengths to constrain \fmu, it may be
difficult to disentangle the contributions by stellar birth clouds and 
the ambient ISM to the total infrared luminosity. Galaxies with cirrus-like
mid-infrared emission in the SINGS sample tend to have low specific star
formation rates and infrared spectra dominated by the emission from 
ambient-ISM dust (as is the case, for example, for NGC~3521 in 
Fig.~\ref{seds2}a). The inclusion of either far-infrared observations to
constrain the temperature of the dust in thermal equilibrium or ultraviolet 
and optical observations to constrain the attenuation of starlight by dust
can lift the ambiguity about the origin of the infrared emission (this is
illustrated by the reasonably tight constraints obtained on \fmu\ for 
NGC~3521 in all the cases considered in Fig.~\ref{pdfs1}). We have checked 
that the predicted mid-infrared emission of the ambient ISM in these 
analyses is always consistent with the expectation that stellar ultraviolet
photons (with $\lambda<3500$~\AA) are the main contributors to the 
excitation of PAH molecules and the stochastic heating of dust grains
responsible for the hot mid-infrared continuum. This is because stars 
slightly older than $10^7$~yr, which have migrated from their birth clouds 
into the ambient ISM, are still bright in the ultraviolet. In the SINGS 
sample, for example, ultraviolet photons generated by stars older than 
$10^7$~yr account for about 5 per cent of the heating of ambient-ISM dust
for galaxies with the lowest specific star formation rates (\ssfr). This
fraction reaches about 85 per cent for galaxies with the highest \ssfr. At 
the same time, PAHs and the hot mid-infrared continuum are found to produce
from about 1 per cent to about 35 per cent, respectively, of the infrared
luminosity of the ambient ISM in these galaxies. Thus, enough stellar 
ultraviolet photons are produced to account for the mid-infrared emission 
from PAHs and hot dust in the ambient ISM of these galaxies, even if a 
large part are absorbed by cooler grains in thermal equilibrium.

\begin{figure}
\centering
\includegraphics[width=0.5\textwidth]{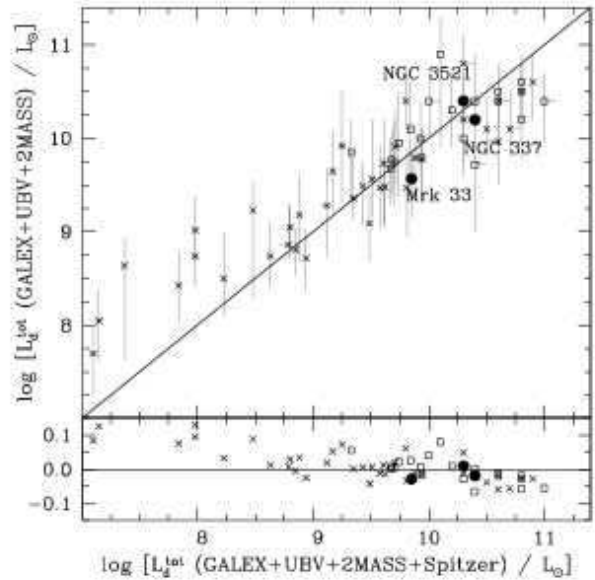}
\caption{Estimates of the total infrared luminosity \ldust\ derived
from fits of the observed ultraviolet ({\it GALEX FUV} and {\it 
NUV}), optical (RC3 {\it UBV}) and near-infrared (2MASS {\it JHKs})
luminosities, plotted against estimates of the same quantity when
including also the constraints from {\it Spitzer} (IRAC 3.6, 4.5,
5.8 and 8.0~\mic\ and MIPS 24, 70, and 160~\mic), for 61 galaxies
with {\it GALEX} measurements in the sample studied in 
Section~\ref{application}.  The different symbols have the same
meaning as in Fig.~\ref{irx_beta}.  The errors bars (in both the
horizontal and vertical directions) represent the 16--84 percentile
range in \ldust\ derived from the likelihood distributions. The 
bottom panel shows the logarithm of the difference
$\mathrm{\ldust({\it GALEX}+{\it UBV}+2MASS)}$ minus 
$\mathrm{\ldust({\it GALEX}+{\it UBV}+2MASS+{\it Spitzer)}}$
divided by $\mathrm{\ldust({\it GALEX}+{\it UBV}+
2MASS+{\it Spitzer)}}$.} 
\label{constraints_ldust}
\end{figure}

The above examples illustrate the usefulness of our model to interpret
multi-wavelength observations of star-forming galaxies, and how particular
observations may be important to constrain specific galaxy parameters.
To further investigate the need for infrared information in the 
determination of \ldust, we compare in Fig.~\ref{constraints_ldust}
estimates of this quantity obtained by fitting {\it GALEX}, RC3 and
2MASS data alone to those obtained when adding all the {\it Spitzer} mid-
and far-infrared constraints (from IRAC and MIPS), for 61 SINGS galaxies
with {\it GALEX} measurements in our sample. For most galaxies, both
estimates are consistent with each other, although the estimates based
on stellar emission alone are far more uncertain than those including
infrared constraints (as can be seen from the much larger vertical than
horizontal error bars in Fig~\ref{constraints_ldust}). There is a tendency
for starlight-based estimates of \ldust\ to lie systematically under
the more precise estimates including infrared constraints for the most
luminous galaxies. This could arise, for example, if the SINGS galaxies
with the largest infrared luminosities contained significant amounts of 
enshrouded stars with little contribution to the emergent optical and 
near-infrared light. At low \ldust, starlight-based estimates of the 
infrared luminosities of blue compact dwarf galaxies in our sample are 
very uncertain in Fig~\ref{constraints_ldust}. We conclude that rough
estimates of \ldust\ may be obtained based on ultraviolet, optical and
near-infrared observations alone (at least for values in the range from 
a few $\times10^{8}$ to a few $\times10^{10}L_\odot$), but reliable
estimates of this parameter require longer-wavelength infrared 
observations.  We note that the total dust luminosities estimated 
including infrared observations are typically within 10 per cent of 
those estimated by \cite{DRAINE07} on the basis of their 
more sophisticated physical dust model.

\subsubsection{Sample statistics}
\label{statistics}

We can estimate the physical parameters of all the 66 SINGS galaxies
in our sample in the same way as exemplified in Figs.~\ref{pdfs1} and
\ref{pdfs2} above for NGC~3521, NGC~337 and Mrk~33. By doing so, we
can explore potential correlations between different parameters of 
star-forming galaxies and infrared colours. In 
Table~\ref{table_correlations}, we present the results of this
investigation. We list the Spearman rank correlation coefficients
for the relations between three {\it Spitzer} infrared colours, 
$\leight/\ltwofour$, $\ltwofour/\lseventy$ and $\lseventy/
\lonesixty$, and 11 parameters derived from our spectral fits of SINGS
galaxies: the specific star formation rate, \ssfr; the fraction of total
infrared luminosity contributed by dust in the ambient ISM, \fmu; the
global contributions (i.e. including stellar birth clouds and the ambient
ISM) by PAHs (\xipahstot), the hot mid-infrared continuum (\ximirtot)
and warm dust in thermal equilibrium (\xiwarmtot) to the total infrared 
luminosity (equations~\ref{exipahstot}, \ref{eximirtot} and 
\ref{exiwarmtot}); the equilibrium temperature of warm dust in stellar birth
clouds, \tbgswarm; the contribution by cold dust in thermal equilibrium
to the total infrared luminosity, \xicoldtot\ (equation~\ref{exicoldtot});
the equilibrium temperature of cold dust in the ambient
ISM, \tbgscold; the star formation rate averaged over the past 100~Myr,
$\psi=M_\ast\psi_S$ (see equation \ref{ssfr_def}); the stellar mass, 
$M_\ast$; and the ratio of dust mass to stellar mass, \mdms\ (see 
equation~\ref{mdust2}). We also indicate in Table~\ref{table_correlations}
the significance levels of these correlations for our sample size of 66
galaxies. 

\begin{table*}
\centering
\caption{Correlations between three {\it Spitzer} infrared colours 
$\leight/\ltwofour$, $\ltwofour/\lseventy$ and $\lseventy/\lonesixty$
and the median likelihood estimates of several physical parameters 
constrained using our model, for the 66 SINGS galaxies in our sample.
For each combination of infrared colour and parameter, the first row
indicates the Spearman rank correlation coefficient $r_S$ of the 
relation between the two quantities, while the second row indicates the
significance level of the correlation for the sample size.}
\begin{tabular}{lrrrrrrrrrrr}
\hline
 & \ssfr \ & \fmu \ & \xipahstot \ & \ximirtot \ & \xiwarmtot \ & \tbgswarm \ & \xicoldtot \ & \tbgscold \ & $\psi$ & $M_\ast$ & \mdms\ \\
\hline
$\leight/\ltwofour$ & $-0.723$ & $0.679$ & $0.715$ & $0.108$ & $-0.858$ & $-0.437$ & $0.663$ & $-0.318$ & $-0.007$ & $0.634$ & $-0.377$ \\
 & $6\sigma$ & $5\sigma$ & $6\sigma$ & $1\sigma$ & $7\sigma$ & $4\sigma$ & $5\sigma$ & $3\sigma$ & $<1\sigma$ & $5\sigma$ & $3\sigma$ \\
\\
$\ltwofour/\lseventy$ & $0.018$ & $-0.201$ & $0.182$ & $0.253$ & $0.212$ & $0.423$ & $-0.390$ & $-0.081$ & $0.503$ & $0.243$ & $0.133$ \\
 & $<1\sigma$ & $2\sigma$ & $1\sigma$ & $2\sigma$ & $2\sigma$ & $3\sigma$ & $3\sigma$ & $<1\sigma$ & $4\sigma$ & $2\sigma$ & $1\sigma$ \\
\\
$\lseventy/\lonesixty$ & $0.466$ & $-0.464$ & $-0.556$ & $-0.162$ & $0.741$ & $0.509$ & $0.577$ & $0.692$ & $-0.123$ & $-0.450$ & $0.040$ \\
 & $4\sigma$ & $4\sigma$ & $4\sigma$ & $1\sigma$ & $6\sigma$ & $4\sigma$ & $5\sigma$ & $6\sigma$ & $1\sigma$ & $4\sigma$ & $<1\sigma$ \\
\hline
\end{tabular}
\\
\label{table_correlations}
\end{table*}

\begin{figure*}
\centering
\includegraphics[angle=270,width=0.75\textwidth]{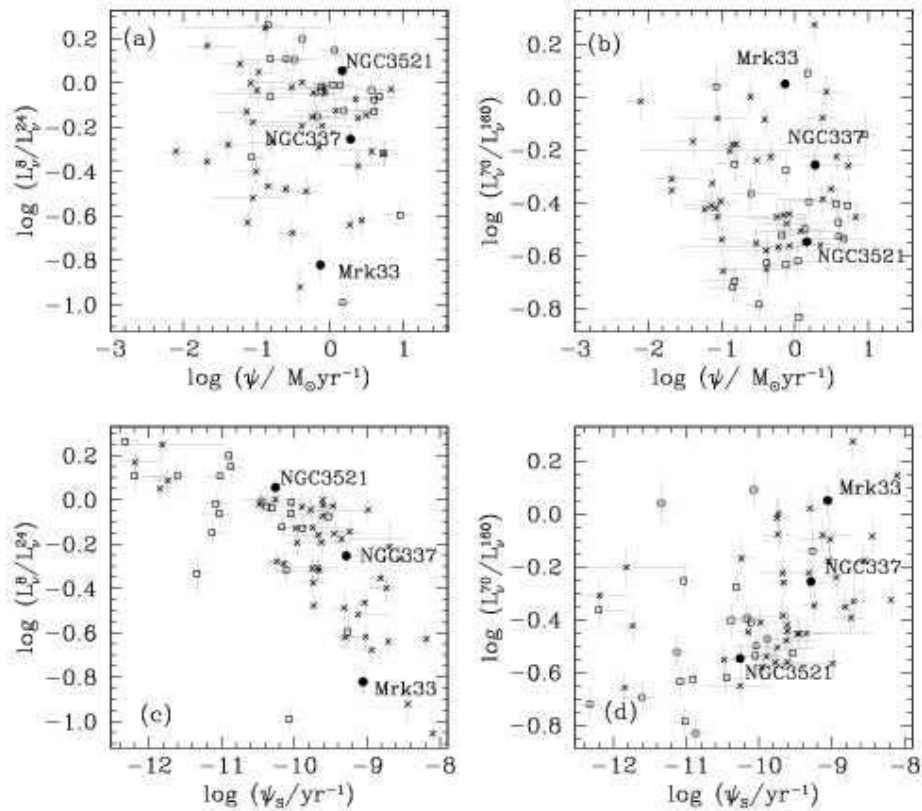}
\caption{{\it Spitzer} infrared colours plotted against median-likelihood
estimates of the star formation rate, for the 66 SINGS galaxies of 
the sample discussed in Section~\ref{application}. (a) Ratio of 8-\mic\ 
to 24-\mic\ luminosity plotted against star formation rate averaged over 
the past 100~Myr, $\psi= M_\ast\psi_S$ (see equation \ref{ssfr_def}). 
(b) Ratio of 70-\mic\ to 160-\mic\ luminosity plotted against $\psi$. 
(c) Ratio of 8-\mic\ to 24-\mic\ luminosity plotted against specific 
star formation rate \ssfr.  (d) Ratio of 70-\mic\ to 160-\mic\ 
luminosity plotted against \ssfr. The different symbols have the same 
meaning as in Fig.~\ref{irx_beta}. For each point, the vertical error
bars represent the $\pm1\sigma$ observational uncertainties, while the
horizontal error bars represent the 16--84 percentile range derived 
from the likelihood distribution of the quantity on the $x$-axis.} 
\label{correlations1}
\end{figure*} 

\begin{figure*}
\centering
\includegraphics[angle=270,width=\textwidth]{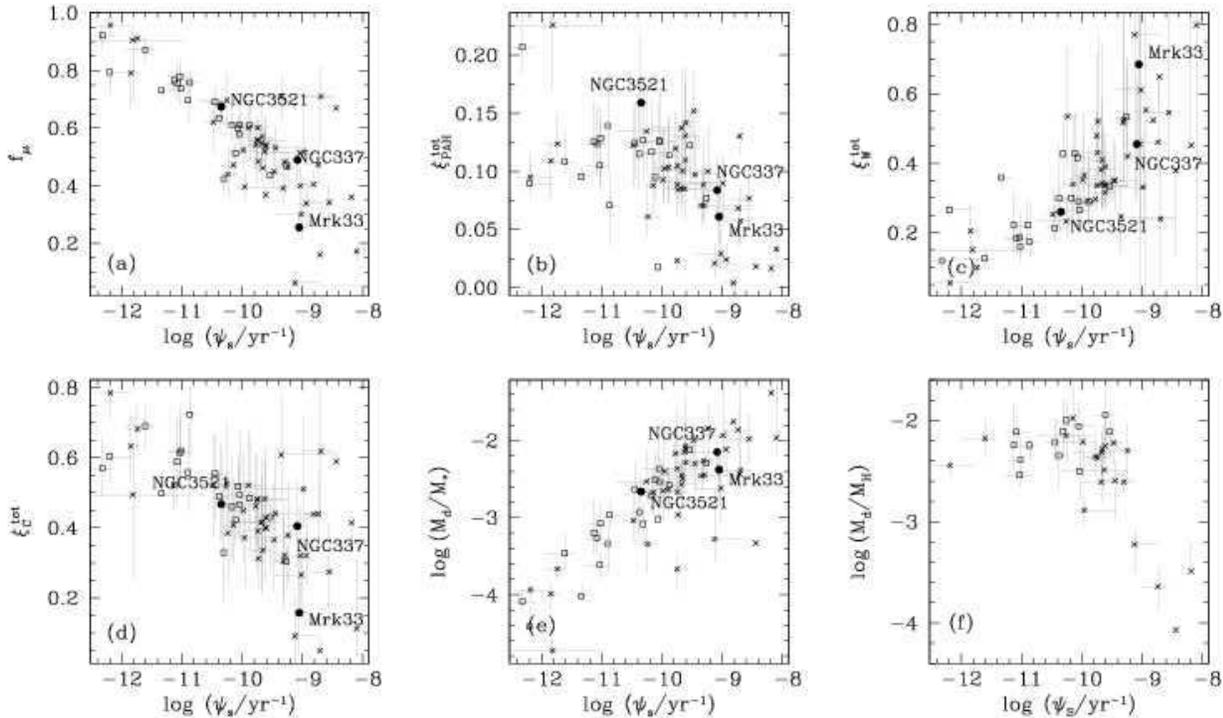}
\caption{Median-likelihood estimates of galaxy properties plotted 
against specific star formation rate, for the 66 SINGS galaxies of
the sample discussed in Section~\ref{application}. (a) Fraction of total
infrared luminosity contributed by dust in the ambient ISM, \fmu. (b) 
Global contribution (i.e. including stellar birth clouds and the ambient
ISM) by PAHs to the total infrared luminosity, \xipahstot. (c) Global
contribution by warm dust in thermal equilibrium to the total infrared
luminosity \xiwarmtot. (d) Contribution by cold dust in thermal 
equilibrium to the total infrared luminosity, \xicoldtot. (e) Ratio 
of dust mass to stellar mass, \mdms. (f) Ratio of dust mass to gas mass,
\mdmg, for the 35 galaxies with gas masses $M_{\mathrm H}=M(\hi+{\mathrm
H}_2)$ available from \protect\cite{DRAINE07}. The different symbols have
the same meaning as in Fig.~\ref{irx_beta}. The error bars represent the
16--84 percentile ranges derived from the likelihood distributions of the 
estimated parameters.}
\label{correlations2}
\end{figure*} 

Table~\ref{table_correlations} shows that the infrared colours
$\leight/\ltwofour$ and $\lseventy/\lonesixty$ correlate well with 
several model quantities, such as \ssfr, \fmu, \xipahstot, \xiwarmtot,
\tbgswarm, \xicoldtot\ and \tbgscold, while $\ltwofour/\lseventy$ does
not. Furthermore, it is interesting to note that the specific star 
formation rate, but {\em not} the star formation rate itself, 
correlates well with $\leight/\ltwofour$ and $\lseventy/\lonesixty$.
This is illustrated in Fig.~\ref{correlations1}, where we show 
$\leight/\ltwofour$ and $\lseventy/\lonesixty$ as a function of both
\ssfr\ and $\psi$ for the 66 galaxies in our sample.  We find that, 
as anticipated in Section~\ref{results}, the correlation between 
infrared colours and specific star formation rate arises from a 
drop in the relative intensity of PAHs and a blueshift of the peak 
infrared luminosity (i.e. a rise in the overall dust temperature) when
the star formation activity increases. 

In Fig.~\ref{correlations2}, we explore the relations between \ssfr\
and other physical properties of the SINGS galaxies, which correlate
well with infrared colours:  \fmu, \xipahstot, \xiwarmtot, \xicoldtot\
and \mdms\ (Table~\ref{table_correlations}). The Spearman rank coefficients
for these correlations are $r_S=-0.764$ (indicating a $6\sigma$ significance
level for the sample size), $-0.551$ ($4\sigma$), $0.739$ ($6\sigma$),
$-0.647$ ($5\sigma$) and $0.780$ ($6\sigma$), respectively. For 
completeness, we also show in Fig.~\ref{correlations2} the relation 
between dust-to-gas mass ratio, \mdmg, and \ssfr, for the 35 galaxies 
with gas masses $M_{\mathrm H}=M(\hi+{\mathrm H}_2)$ available from
\protect\cite{DRAINE07}. In this case, the Spearman rank coefficient
is $r_S=-0.537$, indicating a $3\sigma$ correlation only for the 
reduced sample size. We note that the dust masses in 
Fig.~\ref{correlations2} are typically within 50 per cent of 
those estimated by \cite{DRAINE07} on the basis of their more 
sophisticated physical dust model.

The strong correlations between infrared colours and several physical
quantities of the SINGS galaxies in Fig.~\ref{correlations1} and 
\ref{correlations2} provide important insight into the link between
star formation and ISM properties in galaxies. For example, we also
find a $7\sigma$ correlation between the star formation rate $\psi$ 
and the dust mass $M_{\mathrm d}$ for the 66 SINGS galaxies in our 
full sample, and a $5\sigma$ correlation between $\psi$ and the gas
mass $M_{\mathrm H}$ for the 35 galaxies with \hi\ and ${\mathrm H}_2$
measurements (not shown). To some extent, the rise in the contribution
by stellar birth clouds (i.e. giant molecular clouds) to the infrared
emission from a galaxy when \ssfr\ increases (Fig.~\ref{correlations2}a),
the accompanying rise in the contribution by warm dust 
(Fig.~\ref{correlations2}c), the drop in that by cold dust
(Fig.~\ref{correlations2}d) and the weakening of PAH features
(Fig.~\ref{correlations2}b) have been anticipated in several
previous studies (e.g., \citealt{H86,C96,GRASIL98,DALE01,DALE07}).
The originality of our approach is to {\em quantify} these effects by 
means of a simple but versatile model, which allows statistical studies 
of the star formation and dust properties of large samples of galaxies. 
It is worth emphasising that our method does not introduce these 
relations a priori and that they arise from our consistent treatment 
of stellar populations and dust in galaxies. 

 
\subsection{Comparison with previous models}
\label{comparison_models}

The model we have developed allows one to interpret the infrared spectral
energy distributions of galaxies consistently with the ultraviolet and 
optical emission, in terms of combined constraints on the star formation
and dust properties. At the same time, the model is versatile enough that
it can be used to statistically derive such properties for large samples
of observed galaxies. So far, the tools most widely used to extract 
minimum physical quantities from infrared spectra of large samples of 
observed galaxies are the spectral libraries of \cite{CE01} and 
\cite{DH02}. As an illustration, we compare here the constraints derived
from our model to those that would be obtained using these libraries for
the three galaxies studied in detail in Section~\ref{results}
(Fig.~\ref{seds2}).

\begin{figure}
\centering
\includegraphics[width=0.5\textwidth]{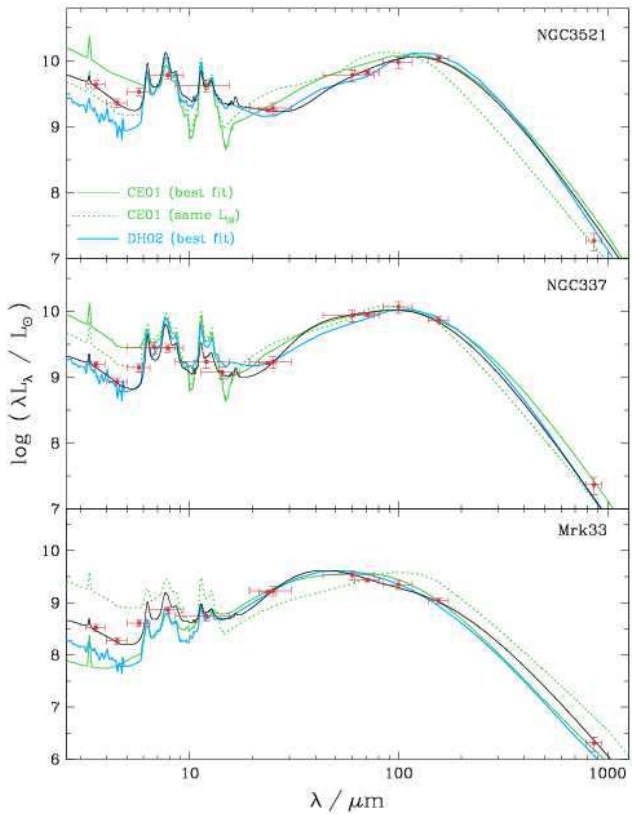}
\caption{Fits by various models to the observed infrared ({\it 
Spitzer} IRAC and MIPS, {\it ISO}, {\it IRAS} and SCUBA) spectral
energy distributions of the same three galaxies as in Fig.~\ref{seds2}
(in red). In each panel, the black line shows the best-fit
model from Fig.~\ref{seds2}; the dotted green line is the 
\protect\cite{CE01} template spectrum corresponding to the same
total infrared luminosity as the black line; the solid green line
is the \protect\cite{CE01} template that best fits the {\it IRAS}
and {\it ISO} observations of the galaxy; and the blue line is 
the \protect\cite{DH02} model that best fits the MIPS observations
of the galaxy (see text for detail).}
\label{comparison}
\end{figure}

\cite{CE01} propose a library of template spectra to reproduce the 
observed correlations between the total infrared luminosity \ldust \ 
($L_\mathrm{IR}$ in their notation) and the luminosities in
individual {\it ISO} and {\it IRAS} bands, for local galaxies. 
This leads them to assign a unique luminosity to a given infrared 
spectral shape. In Fig.~\ref{comparison}, we compare the best-fit
spectral energy distributions inferred from our model for NGC~3521,
NGC~337 and Mrk~33 (in black, from Fig.~\ref{seds2}) with template
spectra from \citet[in green]{CE01}. Since the three galaxies have 
similar infrared luminosities (Fig.~\ref{pdfs2}), they are assigned 
similar infrared spectra in the prescription of \citet[shown
as dotted green lines in Fig.~\ref{comparison}]{CE01}. However, as 
Figs.~\ref{irx_beta} and \ref{seds2} show, NGC~3521, NGC~337 and 
Mrk~33 have different infrared colours. As a result, the \cite{CE01}
template spectra that best fit the {\it ISO} and {\it IRAS}
colours of these galaxies (plotted as solid green lines in 
Fig.~\ref{comparison}) correspond to widely different infrared
luminosities. For example, the \cite{CE01} template that best
fits the observed colours of the dwarf starburst galaxy Mrk~33 
corresponds to a spectral energy distribution typical of an 
ultraluminous infrared galaxy (ULIRG, with $L_\mathrm{IR} \ge 
10^{12} L_{\sun}$). This inconsistency illustrates the impossibility
with the \cite{CE01} library to account for the intrinsic dispersion
of infrared colours among galaxies of comparable infrared luminosity.  

\cite{DH02} parametrise the spectral energy distributions of normal 
star-forming galaxies in terms of a single parameter $\alpha_\mathrm{SED}$,
which is inversely proportional to the intensity of the interstellar
radiation field heating dust grains. Variations in $\alpha_\mathrm{SED}$
can account for the observed range of $F_{60}/F_{100}$ colours of normal
star-forming galaxies. \cite{DALE07} find that the \cite{DH02} models 
that best fit the MIPS colours of NGC~3521, NGC~337 and Mrk~33 have 
$\alpha_\mathrm{SED}=3.32$, 2.23 and 1.44, respectively (these models
are shown as blue lines in Fig.~\ref{comparison}). This sequence in 
$\alpha_\mathrm{SED}$ is consistent with that in the specific star 
formation rate \ssfr\ inferred from our analysis of these galaxies (see 
Fig.~\ref{pdfs1}). As mentioned above, our model can provide greater
insight into the star formation histories and dust properties of galaxies. 

	\subsection{Potential sources of systematic errors}
	\label{systematic_errors}

	\subsubsection{Star formation prior}

We have checked the sensitivity of our results to the prior distributions
of star formation histories in the model library of Section~\ref{approach}.
We have repeated our results using continuous star formations histories 
(i.e. omitting superimposed stochastic starbursts). We find that, in
this case, the overall quality of the fits to the ultraviolet, optical
and infrared spectral energy distributions of SINGS galaxies decreases,
and that the star formation rates of the galaxies are systematically 
lower by up to 40 per cent. We have also tested the influence of the 
initial mass function. Adopting a \cite{SALP55} IMF instead of the 
Chabrier IMF would lead to stellar mass estimates about 1.4 times larger.

	\subsubsection{Attenuation law}
        \label{attenuation_law}

In this study, we have assumed that the effective dust absorption 
optical depth in stellar birth clouds scales with wavelength as
$\hat\tau_\lambda^\mathrm{BC} \propto \lambda^{-1.3}$ (see 
Section~\ref{stellar}). Adopting instead $\hat\tau_\lambda^\mathrm{BC}
\propto \lambda^{-0.7}$ (i.e., the same wavelength dependence as the
attenuation law in the ambient ISM; see \citealt{CF00}) does not change
significantly the overall quality of the fits to the broad-band 
spectral energy distributions of the SINGS galaxies. However, for
the galaxies with spectroscopic observations, the \ha/\hb\ ratios
predicted using $\hat\tau_\lambda^\mathrm{BC} \propto \lambda^{-0.7}$
tend to be lower than observed by typically 15 per cent (see also 
\citealt{WILD07}). This has negligible implications for the constraints
on the dust parameters $\mu$ and \tauv\ derived for the SINGS galaxies.

	\subsubsection{Inclination}

The predictions of our model are averaged over viewing angles. In practice,
however, because of the non-uniform spatial distribution of dust, 
observed fluxes at ultraviolet, optical and infrared wavelengths may 
depend on the angle under which a star-forming galaxy is seen. We account
to some extent for this effect in our model when introducing the uncertainty
$\delta f_\mu$ in the connection between stellar and dust emission
(see Section~\ref{approach}). Also, we do not find any systematic trend
in the quality of spectral fits with the inclination of the SINGS 
galaxies. In the future, we plan to investigate the effect of orientation
further by applying our model to larger samples of galaxies spanning wide
ranges in physical properties and inclinations.


	\subsection{Applicability of the model}
	\label{applicability}
	
Our model is designed primarily to interpret ultraviolet, optical and
infrared observations of large samples of (randomly oriented) galaxies
in terms of effective (i.e galaxy-wide) physical parameters, such as 
star formation rate, stellar mass, dust mass, total infrared luminosity,
breakdown of this luminosity between different dust components and 
between star-forming clouds and the diffuse ISM (Section~\ref{application}).
Constraints on these quantities for statistical samples of galaxies 
at various redshifts are expected to provide useful insight into the 
processes that were important in the evolution of the galaxies we see
today.

Our model is not optimised to interpret in detail multi-wavelength 
observations of individual galaxies, for which the consideration of 
geometrical factors becomes important. Studies of this type generally
require complex radiative transfer calculations for specific optical 
properties of dust grains and specific spatial distributions of stars
and dust. We refer the reader to the more sophisticated models of, for
example, \cite{GRASIL98} and \cite{POPESCU00} for such purposes. Also,
for detailed investigations of the ambient physical conditions of gas
and dust in starburst galaxies, where orientation effects are expected
to be less crucial, the model of \citet[][see also 
\citealt{GROVES07}]{DOPITA05}, which includes radiative transfer in
expanding \hii\ regions, represents a suitable alternative.


	\section{Summary and conclusion}
	\label{conclusion}

We have developed a simple but versatile model to interpret the mid-
and far-infrared spectral energy distributions of galaxies consistently
with the emission at ultraviolet, optical and near-infrared wavelengths.
Our model relies on the \cite{BC03} population synthesis code to compute
the spectral evolution of stellar populations, and on the two-component
model of \cite{CF00} to compute the total infrared luminosity absorbed
and reradiated by dust in stellar birth clouds and in the ambient ISM.
We distribute this total infrared energy in wavelength over the range
from 3 to 1000~\mic\ by considering the contributions by four main dust
components: PAH emission, mid-infrared continuum emission from hot
dust, warm dust with adjustable equilibrium temperature in the range 
30--60~K and cold dust with adjustable equilibrium temperature in the
range 15--25~K. We keep as adjustable parameters the relative 
contributions by PAHs, the hot mid-infrared continuum and warm dust to the
infrared luminosity of stellar birth clouds. Cold dust resides (in 
an adjustabe amount) only in the ambient ISM, where the relative ratios of
the other three components are fixed to the values reproducing the 
observed mid-infrared cirrus emission of the Milky Way. We find that 
this minimum number of components is required to account for the 
infrared spectral energy distributions of galaxies in wide ranges of 
star formation histories.

We have generated a comprehensive library of model galaxy spectra by 
combining a library of attenuated stellar population spectra (built
from stochastic star formation histories and dust contents) with a
library of infrared emission spectra. As Fig.~\ref{seds2} illustrates,
these models provide appropriate fits to the observed ultraviolet, 
optical and infrared spectral energy distributions of nearby galaxies 
in the SINGS sample, for which data are available from {\it GALEX}, RC3,
2MASS, {\it Spitzer}, {\it ISO}, {\it IRAS} and SCUBA \citep{K03}.  We 
have used this model library to derive median-likelihood estimates of the
star formation rate, stellar mass, dust attenuation, dust mass and 
relative contributions by different dust components to the total 
infrared luminosity of every SINGS galaxy in our sample. The accuracy
of these estimates depends on the available spectral information. We
find that, for example, although the total infrared luminosity \ldust\
of a galaxy can be roughly estimated using ultraviolet, optical and 
near-infrared data alone (at least for values in the range from a few
$\times10^{8}$ to a few $\times 10^{10}L_\odot$), reliable estimates 
of this parameter require infrared observations at longer wavelengths.

A main advantage provided by our model is the capacity to study the 
relation between different physical parameters of observed galaxies 
in a quantitative and statistically meaningful way. We find that, 
for example, the specific star formation rate of SINGS galaxies 
correlates strongly not only with observed infrared colours, but 
also with several other properties of the galaxies, such as the 
fraction of total infrared luminosity contributed by dust in the 
ambient ISM, the contributions by PAHs, warm dust and cold dust to
the total infrared luminosity and the ratio of dust mass to stellar
mass. These correlations provide important insight into the link 
between star formation and ISM properties in galaxies. In particular,
they allow one to quantify the relations between star formation rate, 
gas mass, dust mass, stellar mass, dust temperature and distribution
of the dust between giant molecular clouds and the diffuse interstellar
medium. Studies of these relations at different redshifts will have
important implications for our understanding of the physical processes
that dominate galaxy evolution.

Our model should be useful for interpreting data from any modern
galaxy survey at ultraviolet, optical and infrared wavelengths. 
It can also be used to design new observations by optimizing the
set of observables required to constrain specific physical 
parameters of galaxies. This model is meant to be used by the
astronomical community and we intend to make it publicly available.

\section*{Acknowledgments}

We thank the anonymous referee for important suggestions, which helped
improve the quality of this paper. We also thank Brent Groves, Jarle 
Brinchmann and Tim Heckman for comments on the manuscript, and
Jakob Walcher, Armando Gil de Paz, Herv\'e Aussel and H\'el\`ene 
Roussel for helpful discussions. We are grateful to Eli Dwek for 
providing us with an electronic version of the {\it COBE}/FIRAS 
spectrum of the diffuse cirrus emission of the Milky Way. EdC is financed
by the EU Marie Curie Research Training Network MAGPOP. This research
made use of the NASA/IPAC Extragalactic Database (NED) which is operated
by the Jet Propulsion Laboratory, California Institute of Technology,
under contract with the National Aeronautics and Space Administration.

\def\aj{AJ}
\def\araa{ARA\&A}
\def\apj{ApJ}
\def\apjl{ApJ}
\def\apjs{ApJS}
\def\apss{Ap\&SS}
\def\aap{A\&A}
\def\aapr{A\&A~Rev.}
\def\aaps{A\&AS}
\def\mnras{MNRAS}
\def\pasp{PASP}
\def\pasj{PASJ}
\def\qjras{QJRAS}
\def\nat{Nature}
\def\aplett{Astrophys.~Lett.}
\def\aas{AAS}
\let\astap=\aap
\let\apjlett=\apjl
\let\apjsupp=\apjs
\let\applopt=\ao

\bibliographystyle{mn2e}

\bibliography{bib_paperdacunha.bib}

\end{document}